%% file: main.tex
\documentclass[sigconf]{acmart}
\AtBeginDocument{%
  }

\copyrightyear{2026}
\acmYear{2026}
\setcopyright{cc}
\setcctype{by}
\acmConference[ICSE '26]{2026 IEEE/ACM 48th International Conference on Software Engineering}{April 12--18, 2026}{Rio de Janeiro, Brazil}
\acmBooktitle{2026 IEEE/ACM 48th International Conference on Software Engineering (ICSE '26), April 12--18, 2026, Rio de Janeiro, Brazil}
\acmPrice{}
\acmDOI{10.1145/3744916.3773133}
\acmISBN{979-8-4007-2025-3/26/04}

\usepackage{amsmath,amsfonts}
\usepackage[caption=false,font=normalsize,labelfont=sf,textfont=sf]{subfig}
\usepackage{textcomp}
\usepackage{enumitem}
\usepackage{stfloats}
\usepackage{url}
\usepackage{color}
\usepackage{float}
\usepackage{colortbl}
\usepackage{makecell}
\usepackage{multirow}

\newcommand{\xzp}[1]{\textcolor{black}{{#1}}}

\begin{document}

\title{Actionable Warning Is Not Enough: Recommending Valid Actionable Warnings with Weak Supervision}

\author{Zhipeng Xue}
\orcid{0000-0002-9060-4064}
\affiliation{%
  \institution{The State Key Laboratory of Blockchain and Data Security, Zhejiang University}
   \city{Hangzhou}
  \country{China}
}
\email{zhipengxue@zju.edu.cn}

\author{Zhipeng Gao}
\orcid{0000-0003-3030-9917}
\affiliation{%
  \institution{Shanghai Institute for Advanced Study, Zhejiang University}
   \city{Shanghai}
  \country{China}
}
\email{zhipeng.gao@zju.edu.cn}

\author{Tongtong Xu}
\orcid{0000-0002-4323-497X}
\affiliation{%
  \institution{Huawei}
   \city{Hangzhou}
  \country{China}
}
\email{xutongtong9@huawei.com}

\author{Xing Hu}
\orcid{0000-0003-0093-3292}
\affiliation{
  \institution{The State Key Laboratory of Blockchain and Data Security, Zhejiang University}
   \city{Hangzhou}
  \country{China}
}
\email{xinghu@zju.edu.cn}

\author{Xin Xia}
\authornote{Corresponding Author}
\orcid{0000-0002-6302-3256}
\affiliation{
  \institution{The State Key Laboratory of Blockchain and Data Security, Zhejiang University}
   \city{Hangzhou}
  \country{China}
}
\email{xin.xia@acm.org}

\author{Shanping Li}
\orcid{0000-0003-2615-9792}
\affiliation{%
  \institution{The State Key Laboratory of Blockchain and Data Security, Zhejiang University}
   \city{Hangzhou}
  \country{China}
}
\email{shan@zju.edu.cn}


\begin{abstract}
\input{abstract}
\end{abstract}



\maketitle

\section{Introduction}
\label{sec:intro}
\input{intro}

\vspace{-5pt}
\section{Motivation}
\label{sec:moti}
\input{moti}

\section{Approach}
\label{sec:approach}
\input{approach}

\vspace{-10pt}
\section{Evaluation}
\label{sec:eval}

\input{evaluation}

\section{\xzp{Discussion}}
\label{sec:discuss}
\input{discussion}

\section{Related Work}
\label{sec:related work}
\input{relatedwork}

\section{Conclusion}
\label{sec:conclusion}
\input{conclusion}

\begin{acks}
This research is supported by the National Science Foundation of China (No. 62572322).  
This research is partially sponsored by the Shanghai Sailing Program (23YF1446900) and the CCF-Huawei Populus Grove Fund. 
We also thank the anonymous reviewers for their insightful comments and suggestions. 
\end{acks}

\bibliographystyle{ACM-Reference-Format}
\bibliography{main}

\end{document}

%% file: abstract.tex
The use of static analysis tools has gained increasing popularity among developers in the last few years. 
However, the widespread adoption of static analysis tools is hindered by their high false alarm rates (up to 90\%). 
Previous studies have introduced the concept of actionable warnings and built machine-learning method to distinguish actionable warnings from false alarms. 
However, according to our empirical observation, the current assumption used for actionable warning(s) collection is rather shaky and inaccurate, leading to a large number of invalid actionable warnings. 
To address this problem, in this study, we build the first large actionable warning dataset by mining 68,274 reversions from Top-500 GitHub C repositories, we then take one step further by assigning each actionable warning a weak label regarding its likelihood of being a real bug. 
Following that, we propose a two-stage framework called {\sc ACWRecommender} to automatically recommend the actionable warnings with high probability to be real bugs (AWHB). 
Our approach warms up the pre-trained model UniXcoder by identifying actionable warnings task (coarse-grained detection stage) and rerank AWHB to the top by weakly supervised learning (fine-grained reranking stage). 
Experimental results show that our proposed model outperforms several baselines by a large margin in terms of nDCG and MRR for AWHB recommendation. 
Moreover, we ran our tool on 6 randomly selected projects and manually checked the top-ranked warnings from 2,197 reported warnings, we reported top-10 recommended warnings to developers, 27 of them were already confirmed by developers as real bugs. 
Developers can quickly find real bugs among the massive amount of reported warnings, which verifies the practical usage of our tool. 

%% file: intro.tex
The static analysis tools have been utilized for a considerable period of time~\cite{brat2005precise, xue2024selfpico, xue2025clean}. 
These tools are widely used by software developers and companies to detect potential bugs and report warnings in recent years~\cite{Smith2019HowDD, Imtiaz2019ChallengesWR, Tahaei2021SecurityNI, 10.1145/3597503.3639583, gao2024easy, yang2024federated}.
For example, Facebook has developed \textit{Infer}~\cite{Distefano2006ALS, Calcagno2015MovingFW}, a static code analysis tool for checking generic bug patterns (e.g, null pointer exceptions, memory leaks, and race conditions) in their Android and iOS apps (including the main Facebook, Whatsapp, Instagram app, and many others). 
Due to the lightweight analysis and low computational cost, these static analysis tools have gained popularity among developers. 

In spite of the success of static analysis tools, their widespread adoption in various software projects is hindered by two major problems.
First, a high false alarm rate is observed. 
The warnings generated by static analysis tools have a very high false-positive rate (e.g., range up to 90\%)~\cite{HECKMAN2011363}. 
In other words, only a small fraction of the reported warnings are real bugs and/or need to be acted on. 
Second, developers often experience information overloading while using static analysis tools. 
That is, even if the real bugs are reported, developers can become flooded with too many generated warnings. 
They may fail to filter out the false alarms and tend to get lost in massive amounts of irrelevant information.

To fill these gaps and help developers to better make use of static analysis tools, previous researchers have introduced the concept of \textbf{actionable warnings}, namely the warnings that need to be acted on by developers~\cite{Alikhashashneh2018UsingML, Heo2017MachineLearningGuidedSU}. 
Particularly, warnings that are acted on by developers are called actionable warnings. 
In practice, a warning is identified as ``actionable'' if and only if a warning is present in a revision and disappears in a subsequent revision. 
\xzp{More formally, given a sequence of commits $C = {c_1, c_2, \ldots, c_n}$, let $W(i)$ denote the set of warnings reported by a static analysis tool on commit $c_i$ $(1 \leq i \leq n)$.
A warning $w \in W(i)$ is defined as \textit{actionable} if it disappears in any subsequent commit between $c_i$ and $c_n$.}
The software engineering researchers have proposed different techniques for the task of actionable warning identification (AWI)~\cite{wang2018there, Yuksel2013AutomatedCO, Lee2019ClassifyingFP}.

Nonetheless, two significant limitations exist within these studies: Firstly, according to our in-depth analysis, \textbf{the current data collection process of obtaining actionable warnings is inaccurate and unreliable}. 
\xzp{In other words, a considerable proportion of actionable warnings are invalid and not real bugs~\cite{li2024tracking, avgustinov2015tracking}}. 
The use of such uncurated datasets can be a major threat when models are trained with such data. 
Unfortunately, many automatic approaches proposed thus far~\cite{hanam2014finding, heckman2008establishing, Liang2010AutomaticCO} were trained and evaluated on these unreliable actionable warnings, without paying attention to the overall quality of the dataset and the correctness of warning labels. 
Secondly, regarding actionable warning identification, the existing approaches commonly use machine learning (ML) techniques for this task. 
These approaches train ML classifiers by using a set of hand-crafted features to determine whether warnings are actionable or not. 
These features are manually designed by human experts and critically affect the model's performance. 
The warning features determination is \textbf{laborious and error-prone and heavily relies on the domain knowledge of the experts}. 
Moreover, there is no guarantee that these hand-crafted features are well-designed (different studies may design redundant features~\cite{heckman2008establishing, heckman2009model,  Kim2007PrioritizingWC, qiu2021deep}) and helpful (features have little contribution to the warning labels~\cite{wang2018there}). 
Therefore, the research problem we aim to tackle in this paper is:
Given the reported warnings, can we accurately recommend the warnings that are more likely to be real bugs without resorting to manually pre-defined warning features? 


In this study, we aim to rank \textbf{actionable warnings} produced by static analysis tools (\textit{Infer} and \textit{Flawfinder} in this study) and recommend the \textbf{A}ctionable \textbf{W}arning with \textbf{H}igh probability to be real \textbf{B}ug (\textbf{AWHB}). 
To be more specific, we build the first large actionable warnings dataset under weak supervision and propose a two-stage framework, named {\sc \textbf{ACWRecommender}} (\textbf{AC}tionable \textbf{W}arning \textbf{Recommender}), to automate this task. 
Our data collection involves two steps.
First, we collect actionable warnings from the Top 500 popular C projects on GitHub, including 1,889 actionable warnings and 39,052 false alarms.
Second, to identify real bugs from actionable warnings, we propose a weak supervision label strategy where each actionable warning is assigned a weak label using our semantic and structural matching rules.
The label estimates the likelihood of the actionable warning being a real bug.
To recommend AWHB, {\sc ACWRecommender} is mainly divided into two stages, including a coarse-grained warm-up stage and a fine-grained reranking stage. 
In the coarse-grained detection stage, an actionable warnings dataset is used to warm-up a neural network.
In the fine-grained reranking stage, we further fine-tune the model to rank the AWHB to the top by weakly supervised learning~\cite{ma2025alibaba, ma2025swe}. 

To evaluate the effectiveness of our weak supervision label strategy and {\sc ACWRecommender}, we manually check the label accuracy and conduct extensive experiments on the AWHB recommendation task.
By comparing with several baselines, the superiority of our proposed model is demonstrated. 
The experimental results show that: 
1) According to our manual validation, 79.5\% of AWHBs correctly identified real bugs, while only 19.5\% of actionable warnings were true positives that revealed actual issues.
2) Regarding the AWHB recommendation task, our reranker performs better than its three baselines in terms of nDCG and MRR;
3) We further conducted an in-the-wild evaluation, and we report top-10 actionable warnings recommended by our tool to the developers of 6 popular Github projects, 27 of them have been confirmed by developers as real bugs, which further justifies the practical usage of our approach. 
In summary, this work makes the following contributions:

\noindent\textbf{1)} We build a large dataset for checking static analysis tools' actionable warnings from popular GitHub C repositories, which contains 1,889 actionable warnings and 39,052 false alarms generated by \textit{Infer} and \textit{Flawfinder}. Then we propose a weak supervision label strategy to identify \textbf{A}ctionable \textbf{W}arnings with \textbf{H}igh probability to be real \textbf{B}ugs (\textbf{AWHB}). Our manual verification reveals that 81\% of AWHBs are real bugs.
    
    \noindent\textbf{2)} We propose a novel two-stage model, {\sc ACWRecommender} to automate the AWHB recommendation task.
    {\sc ACWRecommender} fine-tuned the large pre-train model UniXcoder with text and code features, avoiding the laborious and error-prone process of designing warning features manually. 
    This work builds on our preliminary demo tool \cite{xue2023acwrecommender} by extending it into a two-stage modeling framework and conducting a comprehensive evaluation to validate its effectiveness and practical value. 
    
    \noindent\textbf{3)} We extensively evaluate our {\sc ACWRecommender} with several baseline methods. Evaluation results show that our model can significantly outperform the baselines in AWHB recommendation. 
    Moreover, we have conducted an in-the-wild evaluation with 6 GitHub projects, we submitted the top-10 reported warnings to the developers, and 27 of them have been confirmed by developers as real bugs. 
    
    \noindent\textbf{4)} We have released our replication package~\cite{replication}, including the dataset and the source code of {\sc ACWRecommender}. 
    As the first attempt for the AWHB recommendation, we hope our research lays a good foundation for follow-up works and facilitates other researchers and practitioners to verify their ideas.

%% file: moti.tex
In this section, we first show several motivating examples of actionable warnings from real-world software projects.

\begin{table}
\centering
    \caption{The Examples of Actionable Warning-fix Commits}
    \small
    \vspace{-10pt}
    \label{Table:The Examples of Actionable Warning-fix Commits}
    \begin{tabular}{|ll|} 
    \hline
      \multicolumn{2}{|l|}{\textbf{EX.1: Repo/Commit:} open62541/d52786e~\cite{open62541/commit/d52786e}}\\
       \hline
        \makebox[0.025\textwidth][l]{367} & \makebox[0.4\textwidth][l]{\textcolor{blue}{static UA\_StatusCode}}\\
         & UA\_NodeMap\_replaceNode(UA\_Node *node) \{\\
          & \quad ...\\
         \rowcolor{red!25} \textbf{373 -} & \quad \textbf{\textcolor{blue}{UA\_NodeMapSlot} *slot = findOccupiedSlot(}\\
         \rowcolor{red!25} \textbf{-} & \quad \textbf{ns, \&node-\textgreater nodeId);}\\
         \rowcolor{green!25} +& \quad \textcolor{blue}{UA\_NodeMapSlot} *slot = findOccupiedSlot(\\
         \rowcolor{green!25} + & \quad ns, \&node-\textgreater head.nodeId);\\

         374 & \quad \textcolor{blue}{if}(!slot))\{\\
         376 & \quad \quad\textcolor{blue}{return} UA\_STATUSCODE\_BADNODEID;\}\\
         \textbf{380} & \quad \textbf{\textcolor{blue}{UA\_NodeMapEntry} *oldEntry = slot-\textgreater entry;} \\

      \hline
      \multicolumn{2}{|l|}{\textbf{Commit Message:}refactor(server): Use a union to avoid cast-}\\
      \multicolumn{2}{|l|}{ing of node classes}\\
    \hline
    \multicolumn{2}{|l|}{\textbf{Warning Type:} Null Dereference }\\
    \hline
      \multicolumn{2}{|l|}{\textbf{Warning Qualifier:} pointer `slot` last assigned on line 373}\\
      \multicolumn{2}{|l|}{could be null and is dereferenced at line 380}\\
  \hline
      \hline
      \multicolumn{2}{|l|}{\textbf{EX.2: Repo/Commit:} libevhtp/d13b72b~\cite{libevhtp/commit/d13b72b}}\\
       \hline
         \textbf{3651} & \quad \textbf{fd = socket(sa-\textgreater sa\_family, SOCK\_STREAM, 0);}\\
         & \quad ...\\
         \rowcolor{green!25} + & \quad \textcolor{blue}{if} (fd != -1)\\
         \rowcolor{green!25} + & \quad \quad evutil\_closesocket(fd);\\
         \textbf{3673} & \quad \textbf{\textcolor{blue}{return} evhtp\_accept\_socket(htp, fd, backlog);}\\
         3674 &\}\\
      \hline
      \multicolumn{2}{|l|}{\textbf{Commit Message:} FIX: Socket leakage on error \#6.}\\
      \multicolumn{2}{|l|}{Cleanup open file descriptors when bind\_sockaddr fails.}\\
    \hline
    \multicolumn{2}{|l|}{\textbf{Warning Type:} Resource Leak }\\
    \hline
      \multicolumn{2}{|l|}{\textbf{Warning Qualifier:} Resource acquired to `fd` by call to }\\
      \multicolumn{2}{|l|}{`socket()` at line 3651 is not released after line 3673.}\\
  \hline
    \end{tabular}
    \vspace{-10pt}
\end{table}

Due to the high false alarm rate of static analysis tools, software engineering researchers have proposed pipelines to collect actionable warnings~\cite{Utture2022StrikingAB, Alikhashashneh2018UsingML, Heo2017MachineLearningGuidedSU}, namely warnings that need to be acted on by developers. 
Upon investigation of these actionable warnings, we have observed that \textbf{actionable warnings collected by the current pipeline are inaccurate and may not necessarily represent real bugs}. This observation is also consistent with the latest empirical findings~\cite{kang2022detecting}.
The underlying reason is the current pipeline regards warnings that exist in one revision and disappear in later revisions as actionable warnings. 
\textbf{However, this assumption is rather shaky because the disappearance of such warnings can be caused by a non-relevant fix/commit}, leading to the introduction of invalid actionable warnings. 
Table~\ref{Table:The Examples of Actionable Warning-fix Commits} demonstrates two actionable warnings, an invalid actionable warning (Ex.1) and a real bug warning addressed by developers (Ex.2).

Ex.1 shows an example of invalid actionable warnings. 
\textit{Infer} reported a \textbf{Null Deference} warning on line 373 because \texttt{slot} could be null and dereferenced. 
However, this warning is invalid because there is a null-checker for \texttt{slot}, and it will never be dereferenced if it is null. 
This warning disappeared in a non-relevant commit and was mistakenly extracted as an actionable warning. 
In contrast, Ex.2 shows a genuine bug warning (\textbf{Resource Leak} at line 3651) reported by \textit{Infer}. 
To obtain a more accurate actionable warning dataset, it is necessary to determine whether the given actionable warning is invalid (i.e., Ex.1) or a genuine bug warning (i.e., Ex.2). 
Upon analyzing the ``disappeared revision'' of the invalid actionable warning and the genuine bug warning, we deduce that the genuine bug warning can be correctly identified by considering two key factors from the fix revision: semantic factor (e.g., commit message) and structural factor (e.g., code change context). 
The commit message conveys the semantic intention of the commit~\cite{liu2018neural, gao2021automating, wang2024makes}, while the code change conveys the syntactic structural information regarding the behavior of the commit~\cite{wang2024just, qiu2021deep}.
For example, the commit message of the genuine bug warning (Ex.2) validates the warning type (e.g., socket leakage), and the code change is a common code pattern for patching resource leakage, suggesting a high probability of this warning being related to a real bug. 
On the contrary, for the invalid actionable warning (Ex.1), the commit message and code change of its disappeared revision have no correlation with the reported \textbf{Null Deference} warning, which implies the likelihood of being a real bug of this warning is relatively low. 

Guided by the motivating example, we propose a weak supervision labeling strategy including semantic (commit message) and syntactic (code change pattern) matching rules to distinguish AWHB from noisy actionable warnings, and use these labels to fine-tune the pre-trained model.

\begin{figure*}
	\centering
    \centerline{\includegraphics[width=0.8\textwidth]{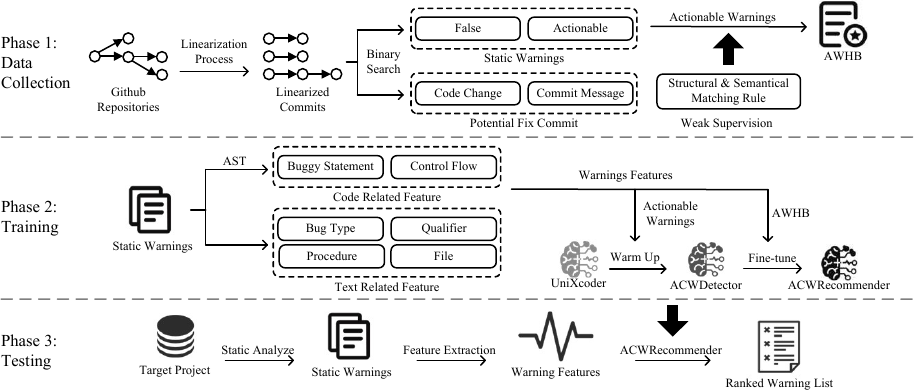}}
    \caption{Overview of Our Approach}
    \label{fig:overview}
\end{figure*}


%% file: approach.tex
The overall framework is illustrated in Fig.~\ref{fig:overview}. 
We present details of our approach design as follows.

\subsection{Actionable Warnings Collection}
The goal of this step is to collect all the actionable warnings from a project and identify potential bug-fix revisions for each actionable warning. 
A commonly used method is to run static analysis tools on the compiled code of each revision and generate a list of warnings for each revision~\cite{liu2019avatar,liu2021mining}.
Then for each warning, check whether it is closed in later revisions (label as actionable) or presented until the last revision (label as a false alarm). 
\xzp{However, the previous method can be time-consuming and resource-intensive, especially for large projects with a long history of revisions.}

\begin{figure*}[htbp]
	\centering
	\includegraphics[scale=0.8]{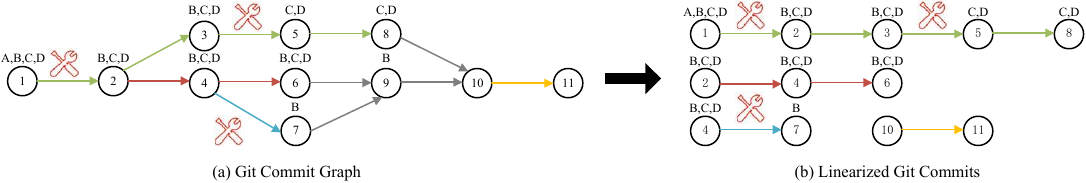}\\
	\caption{Example of the Git Commit Graph}
     \label{Fig:Example of The Git Commit Graph Linearization}
         \vspace{-10pt}
\end{figure*}

\xzp{In this study, we propose a graph-based binary search algorithm to efficiently identify actionable warnings and their corresponding bug-fix revisions, while skipping irrelevant commits during the search process.}
Our graph-based binary search algorithm includes two parts: \texttt{graph linearization} and \texttt{linear binary search}. 
Considering the real-world git histories often contain branches and merges and are organized in graph structures, the \texttt{graph linearization} is responsible for converting a git commit graph into a series of linear git histories. 
Following that, the \texttt{linear binary search} algorithm analyzes each linear git flow to identify actionable warnings and their corresponding bug-fix revisions within this linear git history.

The \texttt{graph lineariztion} algorithm takes a git commit history graph $G$ as input, and outputs all possible linear commit histories within $G$. 
Each node in $G$ represents a specific software reversion, while each edge represents a git commit connecting the previous reversion and its subsequent reversion. 
\xzp{Fig.~\ref{Fig:Example of The Git Commit Graph Linearization} illustrates an example of a Git commit graph where several reported warnings (marked as A, B, C, and D) were fixed and its linear commit histories extracted by our \texttt{graph linearization} algorithm.}
We define \textit{start nodes} as nodes without parent nodes or nodes with more than one child node (such as node 1, node 2, node 4), and \textit{end nodes} as nodes without child nodes (such as node 11). 
\xzp{To perform the linearization, we first remove all the merge commits (e.g., edges marked with gray color) to avoid misleading results, such as falsely attributing a fix to the current branch when it actually originated from another branch via a merge.}
In this context, all the nodes before a merge commit become \textit{end nodes} (such as node 8, node 9 and node 7), and all the nodes after a merge commit become \textit{start nodes} (such as node 10). 
We then obtain all possible linear git commit histories by tracking from a \textit{start node} (including node 1, node 2, node 4, and node 10) until reaching any \textit{end nodes} (including node 6, node 7, node 8 and node 11).  
As illustrated in Fig.~\ref{Fig:Example of The Git Commit Graph Linearization}, the git commit graph will be converted to four linear git commit histories (marked with green/red/blue/yellow color) respectively.

After partitioning the git commit graph into a series of linear git histories, we utilize the \texttt{linear binary search} algorithm to detect actionable warnings and potential fix revisions on each linear git history. 
Specifically, we collect actionable warnings if and only if a warning present in a revision and disappears in its adjacent child revision. 
In this case, the child revision is regarded as a potential bug-fix revision. 
For each linear commit history, we set the \texttt{left pointer} point to the \textit{start node} and the \texttt{right pointer} point to the \textit{end node}. 
If the \texttt{start node} and \texttt{end node} have identical warning lists, then we hypothesize that the reversions between the \textit{start node} and \textit{end node} contain no actionable warning(s) and can be unchecked.
Otherwise, there are actionable warnings between the \textit{start node} and the \textit{end node}. 
If the \textit{start node} is the parent of the \textit{end node}, we collect all warnings that vanished from the \textit{start node} as actionable warnings.
The process continues until all possible actionable warnings of this git linear history are identified. 

Similar to previous research~\cite{wang2018there}, if a warning has not been addressed for more than two years since its first occurrence, it is considered as a false warning.

\vspace{-5pt}
\subsection{Label under Weak Supervision}
\label{subsec:Label under Weak Supervision}
All of the previous research~\cite{Utture2022StrikingAB, Alikhashashneh2018UsingML, Heo2017MachineLearningGuidedSU} focused on detecting actionable warnings only, however, as shown in Fig.~\ref{Table:The Examples of Actionable Warning-fix Commits}(a), the current collected actionable warnings are inaccurate and unreliable. 
\xzp{In this study, we aim to take one more important step by assigning actionable warnings with different probability scores under a weak supervision labeling strategy.}
The higher probability scores indicate actionable warning(s) are more likely to be real bug(s) (referred to \textbf{AWHB} in this study), which should be inspected at the beginning. 
As mentioned in Section~\ref {sec:moti}, the mismatch between warnings and their bug-fix reversions results in \textbf{invalid actionable warnings}.  
To gather more accurate actionable warnings, we estimate the ``matching degree'' of each actionable warning and its bug-fix commit in terms of two perspectives, i.e., semantical matching rule (using commit message) and structural matching rule (using code change context). 
Based on each matching rule, we assign a value to the actionable warning. 
In this study, we focus on four types of warnings (i.e., \textbf{Uninitialized Variable}, \textbf{Resource Leak}, \textbf{Null Dereference}, and \textbf{Dead Store}) reported by the default settings of \textit{Infer} and the \textbf{Buffer Overflow} warning detected by \textit{Flawfinder}. 
\xzp{For each warning type, the first author reviewed 20 warning-fix commits and summarized the most common keywords and fix patterns for this warning type.}

\subsubsection{Semantic Matching Rule.}
The commit message of a reversion typically provides a semantic summary. 
If the bug-fix reversion's commit message relates to the actionable warning semantically, the actionable warning is likely to be a real bug and being fixed by the bug-fix reversion. 
In this step, we extract the commit message from bug-fix reversion and estimate a semantic matching score $CM(x)$ of the actionable warning $x$ by following Equ.\ref{Equ:CM}:
\begin{equation}
  \begin{gathered}
\label{Equ:CM}
   CM(x) = \max\left(\begin{cases}
    3, & \mathbf{WTK} $ in $ m \\
    2, & \mathbf{WCK} $ in $ m \\
    1, & \mathbf{CK} $ in $ m \\
    0, & $No matching$\\
  \end{cases}\right)
  \end{gathered}
\end{equation}
Given an actionable warning $x$, $m$ refers to the commit message of the potential fix revision of $x$.  
We define three types of keywords for semantic matching, namely \textbf{W}arning \textbf{T}ype \textbf{K}eywords ($\mathbf{WTK}$), \textbf{W}arning \textbf{C}ontext \textbf{K}eywords ($\mathbf{WCK}$), and \textbf{C}ommon \textbf{K}eywords ($\mathbf{CK}$). 
\textbf{WTK} includes the keyword related to the specific warning type. 
\textbf{WCK} includes the critical phrase in the warning qualifier template. 
\textbf{C}ommon \textbf{K}eyword (CK) includes the common word about warning fixing, as shown in Table~\ref{Table:Semantic Matching Rule}. 
In particular, we perform our semantic matching as follows: 
(a) Specifically, if the commit message $m$ contains words matching the warning type keywords (e.g., initial, resource, leak), we consider this actionable warning \textbf{highly likely} to be a real bug, and assign it a semantic score of 3. 
(b) The second column of Table~\ref{Table:Semantic Matching Rule} shows the warning qualifier template of each warning type, we extract the context keyword from the qualifier template as shown in the fourth column. 
If such warning context keywords are mentioned in the commit message $m$, we assume the actionable warning is \textbf{likely} to be a real bug  
and assign it a semantic score of 2. 
(c) We assign a semantic score of 1 to an actionable warning when commit message $m$ only contains fix-related common keywords (e.g., fix, repair, bug), because the fixed bug \textbf{may not} match the corresponding actionable warning. 
(d) In the last, if the commit message $m$ does not match any keyword in the matching rule, it is \textbf{unlikely} the actionable warning relates to a real bug, we assign it a semantic score of 0. 
Especially, if the commit message $m$ matches several rules at the same time, we retain the maximum value of the semantic matching score.

\begin{table*}
  \begin{center}
  \small
    \caption{Semantic Matching Rule}
    \vspace{-10pt}
    \label{Table:Semantic Matching Rule}
    \begin{tabular}{|c|c|c|c|c|} 
      \hline
        Warning Type & Warning Qualifier Template & Warning Type Keyword & Warning Context Keyword& Common Keyword\\
        \hline
        \hline
        Uninitialized Variable& \makecell{The value read from \textit{variable} \\was never initialized} & \makecell{initial, define, \\assign, declare}& \textit{variable} &\multirow{5}*{\makecell[c]{\\\\fix, repair, bug, \\warning, solve, \\problem, handle,\\eliminate, address, \\issue, fail,\\error, exception,\\patch, crash}}\\
        \cline{0-3}
        Null Dereference&\makecell{\textit{pointer} last assigned on line \# could \\be null and is dereferenced at line \# }& \makecell{dereference,\\null pointer, null check,\\NullPointerException}&\textit{pointer} &\\
        \cline{0-3}
        Resource Leak &\makecell{Resource acquired to \textit{variable} \\by call to \textit{function} at line \# \\is not released after line \#} &\makecell{resource, leak, release,\\ cleanup, alloc, clear,\\close, free, destroy,\\terminate, end} &\textit{variable}, \textit{function} &\\
        \cline{0-3}
        Dead Store& \makecell{The value written to\\ \textit{variable} is never used} & \makecell{dead store, \\unused, redundant}& \textit{variable}&\\
        \cline{0-3}
        Buffer Overflow& \makecell{without a limit specification,\\ \textit{function} permits buffer overflows} & buffer, overflow  & \textit{function} & \\
      \hline
    \end{tabular}
  \end{center}
\end{table*}

\subsubsection{Structural Matching Rule.}
Besides using semantic information, we also incorporate structural information (i.e., code change context) to determine if an actionable warning is fixed by the code change of the bug-fix reversion correspondingly. 
Similarly, for each actionable warning, we obtain the code changes from the corresponding potential bug-fix commit and assign a structural matching score based on our code change matching rules.
A high score indicates the code change is likely to fix the actionable warning, and the actionable warning is likely to be a real bug. 
We first manually summarize the common fix pattern for each warning type, as listed in the second column in Table~\ref{Table:Structural Matching Rule}. If the bug-fix reversion's code change matches the fix pattern of a specific warning type, the code change is likely to fix the actionable warning and we assign a structural matching score $CC(x)$ to this actionable warning $x$ by following Equ.~\ref{Equ:CC}.
\begin{equation}
  \begin{gathered}
\label{Equ:CC}
   CC(x) = \max\left(\begin{cases}
    2, & c $ match $ \textbf{Fix Pattern} \\
    1, & c $ match $ \textbf{Scope Pattern} \\
    0, & $No matching$\\
  \end{cases}\right)
  \end{gathered}
\end{equation}

\xzp{
Given an actionable warning $x$, c refers to the code change of the potential fix revision of $x$. 
We define two types of patterns of structural matching, namely fix pattern and scope pattern. 
Fix pattern refers to our summarised bug-fix patterns for each warning type, and scope pattern refers to the expected location of the bug-fix code change.
In particular: (a) If the commit code change matches the summarized fix pattern in the second column of Table~\ref{Table:Structural Matching Rule}, we consider this code change is responsible for resolving the corresponding warnings, and assign a structural matching score of 2 to this actionable warning to indicate the actionable warning is very likely to be fixed by its associated bug-fix reversion.
(b) In addition to using the fix pattern as the strict checking rule, we also consider the scope pattern as a loosened standard for our study. 
The scope pattern refers to the expected location of the bug-fix code change (shown in the last column of Table~\ref{Table:Structural Matching Rule}). 
If the code changes fail to match the fix pattern while falling within the expected bug-fix scope, we consider the actionable warning as a potential bug and assign a score of 1. 
(c) For the actionable warning that does not match any structural matching rule, we assign it a score of 0.
}

\begin{table}
  \begin{center}
  \small
    \caption{Structural Matching Rule}
    \label{Table:Structural Matching Rule}
    \begin{tabular}{|c|c|c|} 
  \hline
    Warning Type & Fix Pattern & Scope Pattern \\
    \hline
  \hline
    Uninitialized Variable & \makecell{assign value by \\assignment or reference} & before warning\\
    \hline
    Null Dereference & add a null-check & before warning\\
    \hline
    Resource Leak & \makecell{invoke resource \\free-related function} & after warning\\
    \hline
    Dead Store & \makecell{use the variable,\\remove assignment} & after warning\\
      \hline
    Buffer Overflow & \makecell{use the safe function,\\add a boundary check} & before warning\\
          \hline
    \end{tabular}
  \end{center}
\end{table}

\subsubsection{Aggregation of Weak Labels}
To obtain a more reliable and robust label for each actionable warning, 
we use majority voting to combine the above semantic matching score and structural matching score, as demonstrated in Equ.~\ref{Equ:Label Aggregation Eqution}.

\begin{equation}
  \begin{gathered}
\label{Equ:Label Aggregation Eqution}
  Label(x) = \begin{cases}
    {\bf \texttt{VTB}}, & CM(x)+CC(x)>3 \\
    {\bf \texttt{LTB}}, & 2<=CM(x)+CC(x)<=3 \\
    {\bf \texttt{UTB}}, & 0<=CM(x)+CC(x)<2 \\
  \end{cases}
  \end{gathered}
\end{equation}

Given an actionable warning $x$, $CM(x)$ refers to the semantic score for $x$ using the commit message matching rule, and $CC(x)$ refers to the structural score for $x$ using the code change matching rule. 
(a) If the sum of $CM(x)$ and $CC(x)$ is greater than 3, it means the actionable warning is very likely to be a real bug from both semantic and structural aspects and we label $x$ as {\bf \texttt{VTB} (Very Likely To be Bugs)}. 
For example, when the commit message matches the bug type (i.e., $CM(x)=3$) and the code change matches the fix pattern (i.e., $CC(x)=2$) at the same time, we can almost ensure the actionable warning $x$ is a real bug. 
(b) Similarly, if the sum of $CM(x)$ and $CC(x)$ falls between 2 and 3, it means the warning $x$ matches the bug-fix reversion from either the semantic or structural aspect and is likely to be a real bug, we label $x$ as {\bf \texttt{LTB} (Likely To be Bugs)}. 
(c) Lastly, if the sum of $CM(x)$ and $CC(x)$ is less than 2, it means the actionable warning $x$ mismatches the bug-fix reversion and is unlikely to be a real bug, we label such instances as {\bf \texttt{UTB} (Unlikely To be Bugs)}. 

So far, we assign each actionable warning with a weak label representing its likelihood of being a real bug. 
We then define the warning $x$ whose $Label(x)$ is {\bf \texttt{VTB}} or {\bf \texttt{LTB}} as \textbf{AWHB} (\textbf{A}ctionable \textbf{W}arning with \textbf{H}igh probability to be real \textbf{B}ug). In other words, actionable warnings that are likely or very likely to be real bugs are referred to as \textbf{AWHB} in this study, which is defined as follows: 
\begin{equation}
  \begin{gathered}
\label{Equ:AWHB define}
\textbf{AWHB} = \{x | Label(x) \in \{\texttt{VTB}, \texttt{LTB}\}\}
  \end{gathered}
\end{equation}

\subsection{Two-stage Model}
So far, we have collected actionable warnings and further labeled the AWHB under weak supervision. 
Due to the data imbalance between the false positive warnings, actionable warnings and AWHB, we propose a two-stage model which includes the coarse-grained detection stage and fine-grained reranking stage.
In the coarse-grained detection stage, the actionable warnings dataset is used to warm-up and let the model learn how to distinguish actionable warnings from false alarms and acquire a basic semantic understanding of static analysis warnings.
In the fine-grained reranking stage, we further fine-tune the model to rerank the AWHB to the top by weakly supervised learning.

\subsubsection{Warm-up the Detector Model}
At this stage, the model is warmed up by learning to distinguish actionable warnings from false alarms, which helps it acquire a fundamental semantic understanding of static analysis warnings.

For a given reported warning, two types of key information are prepared as input for UniXcoder, i.e., text related input (e.g., bug type) and code related input (e.g., AST). 
UniXcoder is well-suited for our encoding task because it combines multi-modal data (comment and AST) during training.
In particular, regarding the text related input, we extract the bug type, qualifier, procedure, and filename from warning reports. Bug type refers to the category of software issue, qualifier provides additional information, procedure outlines the specific function warning occurred, and filename identifies the location of the bug in the code.
Regarding the code related input, we pinpoint the bug location of the reported warning and extract the associated buggy statement for this warning. We then check the parent node of the buggy statement from AST and add control flow information of the buggy statement from AST as code context. 
We feed the text related input and code related input into UniXcoder and obtain the embedding by directly concatenating the outputs of the pre-trained model. 


Warm-up is a crucial step in the fine-tuning process of the pre-trained model, as it helps to stabilize the model's weights, reduce overfitting, and allow it to adapt to the new task more effectively~\cite{Song2022AdaptiveRS}.
In this work, we began by warming up the pre-trained model UniXcoder through identifying actionable warnings. 
The actionable warning identification task can be viewed as a binary classification problem. 
That is, for a given reported warning $x$, we use the model $f(x; \boldsymbol{\theta})$ to determine whether $x$ is actionable or not. 
The actionable warning dataset without weak supervision is used to warm up $f(x; \boldsymbol{\theta})$ and the optimization goal is defined as follows:

\begin{equation}
  \begin{gathered}
\min _{\boldsymbol{\theta}} \frac{1}{N} \sum_{x \in \mathcal{X}} \mathcal{L}\left(f(x ; \boldsymbol{\theta}), {y}_{x}\right)
  \end{gathered}
\end{equation}

where $\mathcal{L}$ denotes a loss function and ${y}_{x}$ is the actionable warning label without weak supervision. Any loss function suitable for a classification task can be used in the warm-up process, and in this study, we use the Binary Cross Entropy Loss.

\subsubsection{Fine-tune the Reranker Model}
To rank different levels of actionable warnings, we transform the ranking problem into a multiclass classification task.
That is, for a given reported warning $x$, we aim to predict $x$ as \texttt{V}/\texttt{L}/\texttt{U}/\texttt{False Warning} based on our actionable warnings dataset under weak supervision. 
The optimization problem is defined as follows:

\begin{equation}
  \begin{gathered}
\min _{\boldsymbol{\theta}} \frac{1}{N} \sum_{x \in \mathcal{X}} \mathcal{J}\left(g(f(x ; \boldsymbol{\theta})), \tilde{y}_{x}\right)
  \end{gathered}
\end{equation}

The $\tilde{y}_{x}$ is the label aggregated from weak supervision. $g(x)$ is the softmax function to compute the probability of each class for $x$, and $\mathcal{J}$ is Cross Entropy Loss, which is suitable for the multiclass classification task. 
When the optimization is done, the final ranking score for each warning $x$ can be inferred as follows:

\begin{equation}
  \begin{gathered}
\label{Equ:Ranking Score}
  \mathcal{S}(x) = \begin{cases}
    class(x) + g_{\tilde{y}}(x), & \overline{y} \in \texttt{VTB}, \texttt{LTB}, \texttt{UTB} \\
    class(x) - g_{\tilde{y}}(x), & \overline{y} \in \texttt{False Warning} \\
  \end{cases}
  \end{gathered}
\end{equation}


where $class(x)$ maps each warning $x$ to a base class score 0/1/2/3 if the predicted class is \texttt{False Warning}/\texttt{UTB}/\texttt{LTB}/\texttt{VTB}, $\overline{y}$ denotes the predicted class of $x$ and $g_{\tilde{y}}(x)$ denotes the probability of the predicted class. 
Suppose $x_{1}$ is predicted as \texttt{VTB} with a probability of 0.6, then the final ranking score $\mathcal{S}(x_{1})=3.6$; 
if $x_{2}$ is predicted as \texttt{False Warning} with a probability of 0.7, then its final ranking score $\mathcal{S}(x_{2})=-0.7$. 
Finally, all the actionable warnings are ranked by their final ranking scores for recommendation. 




%% file: evaluation.tex
\subsection{Data Preparation}
We build our actionable warning dataset by collecting data from the top 500 repositories (ordered by the number of stars) in GitHub for C repositories. 
To the best of our knowledge, this is the first actionable warning dataset collected by mining the histories of popular GitHub C repositories. 
In this study, we adopt two static analysis tools, i.e., \textit{Infer} and \textit{Flawfinder}, to detect potential warnings in software systems. 
\xzp{For \textit{Infer}, we use the default settings to generate warnings on our dataset. In total, Infer reported 31,380 warnings, among which 99.1\% (31,098 warnings) belong to the top four warning types: \textbf{Uninitialized Variable}, \textbf{Resource Leak}, \textbf{Null Dereference}, and \textbf{Dead Store}. Therefore, we focus on these four categories in our study.}
For the \textit{Flawfinder}, we set the priority level to 4, which indicates the vulnerability with high risks, and collect the \textbf{Buffer Overflow} type warnings.

To save time and resources, we have implemented a filtering process for our projects. 
First, we exclude any projects that require more than half an hour to compile and test using SA tools. 
Additionally, we filter out projects whose latest revisions cannot be compiled successfully. 
Finally, we collect both actionable and false warnings from 68,274 revisions of the 394 projects. 
To avoid the double counts of warnings, following the setting of \textit{Infer}, we deduplicated warnings according to the unique hash value of each warning, which is generated by bug type, located file, located procedure, and context code. 
Moreover, by comparing the reported bug location, we make sure there are no warnings duplicated reported by \textit{Infer} and \textit{Flawfinder}.
The detail of the collected warning dataset is listed in Table~\ref{Table:Data Statistics}.

Table~\ref{Table:Data with Weak Supervison Statistics} illustrates the commit message and weak supervision label of the collected 1,889 actionable warnings and the aggregated labels of them based on the Equ.~\ref{Equ:Label Aggregation Eqution}. 
Following the settings of previous work~\cite{wang2018there, yang2021learning, yang2021understanding}, the current existing warnings that haven't been acted by developers for more than two years are regarded as false warnings. 
As a result, 39,052 warnings in our study are assigned with labels of {\bf \texttt{False Warnings}}.
We regard the actionable warning whose aggregated labels are {\bf \texttt{VTB}} (Very Likely To be Bugs) and {\bf \texttt{LTB}} (Likely To be Bugs) as \textbf{AWHB}. 
As can be seen, AWHB only makes up a small proportion (287/40,941) of the total reported warnings.

\xzp{To address the data imbalance in our dataset, in the warm-up stage, we intentionally preserved the original distribution of warnings to allow the model to learn the general characteristics of both actionable and false warnings. In the fine-tuning stage, we first oversampled the actionable warnings (10×) to balance the training dataset. Then, we employed a weighted cross-entropy loss, where each actionable warning was assigned a weight based on its weakly supervised labeled score, which helps the model pay more attention to the AWHB.}

\begin{table}
  \begin{center}
    \caption{Data Statistics}
        \vspace{-10pt}
    \label{Table:Data Statistics}
    \begin{tabular}{|c|c|c|}
  \hline
    Warning Label & Warning Type & Count\\
    \hline
    \hline
    \multirow{6}{*}{Actionable Warning} &Uninitialized Variable & 164  \\
    \cline{2-3}
                                        &Null Dereference & 238 \\
    \cline{2-3}
                                        &Resource Leak & 50  \\
    \cline{2-3}
                                        &Dead Store &  338 \\
    \cline{2-3}
                                        &Buffer Overflow &  1,099 \\
    \cline{2-3}
                                        &Total Warning & 1,889  \\
    \hline
    \multirow{6}{*}{False Warning} &Uninitialized Variable &  7,660 \\
    \cline{2-3}
                                        &Null Dereference & 10,544 \\
    \cline{2-3}
                                        &Resource Leak & 842  \\
    \cline{2-3}
                                        &Dead Store & 11,544  \\
    \cline{2-3}
                                        &Buffer Overflow & 8,462  \\
    \cline{2-3}
                                        &Total Warning & 39,052  \\
    \hline
    \end{tabular}
  \end{center}
      \vspace{-10pt}
\end{table}

\begin{table}
  \begin{center}
    \caption{Data with Weak Supervision Statistics}
        \vspace{-10pt}
    \label{Table:Data with Weak Supervison Statistics}
    \begin{tabular}{|c|c|c|} 
      \hline
    Warning Label & Warning Type & Count\\
    \hline
    \hline
    \multirow{3}{*}{\makecell{Commit Message\\ Matching Rule}} &Warning Type Keyword & 143  \\
    \cline{2-3}
                                        &Warning Context Keyword & 44 \\
    \cline{2-3}
                                        &Common Keyword & 550  \\
    \cline{2-3}
                                        &No Matching & 1,152  \\
    \hline
    \multirow{3}{*}{ \makecell{Code Change\\ Matching Rule}} &Patch Pattern & 228  \\
    \cline{2-3}
                                        &Scope Patter & 396 \\
    \cline{2-3}
                                        &No Matching &  1265 \\
    \hline
        \multirow{4}{*}{\makecell{Aggreated Lablel}} & \texttt{VTB} &  151 \\
    \cline{2-3}
                                        & \texttt{LTB}  & 136 \\
    \cline{2-3}
                                        & \texttt{UTB}  & 1,602  \\
        \cline{2-3}
                                        & \texttt{False Warning} &  39,052 \\
    \hline
    \end{tabular}
  \end{center}
\end{table}


\subsection{Baselines}
To demonstrate the effectiveness of our proposed model, {\sc ACWRecommender}, we compared it to the following selected baselines:

\noindent\textbf{Golden Feature-RF:} 
    The golden feature~\cite{wang2018there} is currently the state-of-the-art approach for actionable warning identification, which applies a random forest classifier to the 23 most influential features.
    We followed the experimental settings of it after removing 5 features that may introduce data leakage~\cite{kang2022detecting}.

\noindent\textbf{Golden Feature-SVM:} 
    Further research~\cite{yang2021learning} on the Golden Features of Wang et al.~\cite{wang2018there} showed that a linear SVM algorithm was an optimal choice for identifying actionable warnings.

\noindent\textbf{Random Forest:} 
    According to our preliminary experiment among more than 10 machine learning models based on the warning feature embedding from UniXcoder, we selected Random Forest as a baseline, which performs best in actionable warning identification.

\noindent \textbf{Random Ranking:} In the AWHB recommendation task, the Random ranking strategy shuffles the warning dataset randomly.

\subsection{Evaluation Metrics}

To evaluate the performance of AWHB recommendation, we use the metrics including MRR and nDCG@K, which are widely-used metrics in information retrieval and recommender system~\cite{Gao2022IKW, gao2020technical, su2021reducing, qiu2021deep}. Mean Reciprocal Rank (MRR) measures how quickly the first AWHB is found in the recommended warning list.
The normalized discounted cumulative gain at K (nDCG@K) measures the quality of the recommendation by comparing the recommended warnings list to a ground truth warnings list. nDCG@K takes into account both the relevance of AWHB and its position in the recommended warnings list. 
A higher MRR and nDCG@K value indicates a better recommendation result.

\subsection{Quantitative Analysis}
\subsubsection{RQ1: The Effectiveness of Weak Supervised Labeling.}
\label{rq1}

\begin{table*}[htbp]
  \begin{center}
    \caption{Labeling Effectiveness Evaluation}
        \vspace{-10pt}
    \label{Table:Labeling Effectiveness Evaluation}
    \begin{tabular}{|c|l|c|c|c|c|c|c|} 
  \hline
    \multicolumn{2}{|c|}{Warning Label} & Uninitialized Variable &Null Dereference & Resource Leak & Dead Store&Buffer Overflow& Total\\
    \hline
    \hline
    \multirow{3}{*}{\makecell{Actionable\\ Warning}} & \# of Samples & 28 & 35 & 14  & 51 & 72 & 200\\
    \cline{2-8}
     & \# of Real Bugs & 5 & 10 & 2  & 7 & 15& 39 \\
         \cline{2-8}
     & Real Bugs Ratio & 5\slash 28 & 10\slash 35 & 2\slash 14& 7\slash 51 & 15\slash 72 &19.5\% \\
     \hline
     \multirow{3}{*}{\makecell{AWHB}} & \# of Samples & 28 & 35 & 14  & 51 & 72 & 200\\
         \cline{2-8}
     & \# of Real Bugs & 25 & 31 & 11  & 29 & 63 & 159 \\
         \cline{2-8}
     & Real Bugs Ratio & 25\slash 28 & 31\slash 35 & 11\slash 14& 29\slash 51 & 63\slash 72 & 79.5\%  \\
     \hline
    \end{tabular}
  \end{center}
\end{table*}

To evaluate the effectiveness of our weak supervision labeling strategy in identifying real bugs, we manually inspected a sample of the actionable warnings and our labeled data respectively. 
Specifically, we randomly sampled 200 actionable warnings and AWHBs from our study, respectively. 
We then invited three software developers, each with at least five years of C programming experience, to review each sample and identify if the warning is valid and represents a real bug.
\xzp{All developers were asked to independently review each warning, following clear instructions that clarified the evaluation criteria: \textit{Please determine whether the given commit fixed the warning reported by the static analysis tool based on: (1) the warning report; (2) the warning’s surrounding code context; (3) the related code changes; and (4) the commit messages.}}
\xzp{To minimize potential order effects, all warnings were presented to reviewers in a randomized order. 
Additionally, no additional information such as CWE labels or model predictions, was provided to avoid confirmation bias.}
After collecting results from each developer, the first author engaged in and had a discussion with three developers when different opinions were met. 
Finally, each warning is assigned with a unique label (i.e., real bug or false alarm) after discussions. 
The results of the manual validation are shown in Table \ref{Table:Labeling Effectiveness Evaluation}:

\textbf{Only 19.5\% of actionable warnings represent real bugs}, validating our empirical findings in Sec.\ref{sec:moti} that the actionable warning identified by prior assumption may not necessarily indicate real issues. 
    An example is shown in Table~\ref{Table:The Examples of Actionable Warning-fix Commits} Ex.1, 
    \textit{Infer} falsely reported a \textbf{Null Dereference} warning in a previous revision and this warning was recognized as an actionable warning. 
    According to previous empirical studies~\cite{HECKMAN2011363}, 35\% to 91\% warnings reported by SA tools are spurious false alarms. 

\textbf{Our weak supervised labeling method is effective for identifying real bugs, 79.5\% of AWHBs are verified as true bugs by developers.} 
    For example, 25 out of 28 \textbf{Uninitialized Variable} warnings are validated as realistic bugs. 
    The matching from semantic and structural aspects allows our weak supervised labeling process to accurately identify a real \textbf{Uninitialized Variable} bug.

\textbf{Our weak supervised labeling method performs consistently well for different SA tools and warning types}. 
    For example, in terms of \textit{Infer}, the real bug ratio of four warning types (\textbf{Uninitialized Variable}, \textbf{Null Dereference}, \textbf{Resource Leak}, and \textbf{Dead Store}) is 75\% (96\slash 128), in terms of \textit{Flawfinder}, 
    the \textbf{Buffer Overflow} warning real bug ratio is 87.5\% (63\slash 72), which shows the robustness of our method.

\subsubsection{RQ2: The Recommendation Effectiveness Evaluation}
Considering that more than 80\% of AWHBs are validated as real bugs in RQ1, we regard the AWHBs as warnings that should be equally recommended and handled earlier. 
In this research question, we evaluate the AWHB recommendation effectiveness of our fine-tuned reranker model. 
During the evaluation, we built 100 test samples by randomly selecting 1,000 warnings from the testing set. 
For each test sample, we ensured there were at least 5 AWHBs within the 1,000 sampled warnings (considering the actionable warning ratio is around 1:20), each test sample was then shuffled for evaluation. 
For each test sample, we aim to evaluate how often the AWHBs are ranked higher among other warnings. 
Thus, the widely-accepted metrics nDCG@K and MRR are used to measure different methods' ranking performance. 
In particular, regarding baseline methods, we use the actionable warnings' probability scores outputted by the baseline methods for ranking. 
Regarding our proposed model, we use the final ranking score of {\sc \textbf{ACWRecommender}} (defined in Equ~\ref{Equ:Ranking Score}) for ranking. 
The evaluation result is listed in Table~\ref{Table:Recommendation Effectiveness Evaluation}. From the table, we can observe the following points:

\begin{table}
  \begin{center}
    \caption{Recommendation Effectiveness Evaluation}
    \vspace{-10pt}
    \small
    \label{Table:Recommendation Effectiveness Evaluation}
    \begin{tabular}{|c|c|c|c|c|} 
\hline
    Measure & nDCG@1 & nDCG@3 & nDCG@5 & MRR \\
    \hline
    \hline
    Random Ranking & 0.020 & 0.031 & 0.052 & 0.007\\
    \hline
        Golden Feature-SVM & 0.074 & 0.095 & 0.137 & 0.064\\
    \hline
    Golden Feature-RF & 0.087 & 0.104 & 0.152 & 0.121\\
    \hline
    Random Forest & 0.143 & 0.211 & 0.239 & 0.204 \\
    \hline
{\sc \textbf{ACWRecommender}} & \textbf{0.356}& \textbf{0.426} & \textbf{0.477}  &\textbf{0.452}\\
    \hline
    \end{tabular}
  \end{center}
\end{table}

 Although initially designed for identifying actionable warnings, we attempted to rank the warnings based on their score from Golden Feature-based approaches and utilized it as a baseline method in recommendation evaluation. 
 Similar to the actionable warning identification, the Golden Feature-based approaches (i.e., \textbf{Golden Feature-SVM} and \textbf{Golden Feature-RF}) outperform \textbf{Random Ranking} in recommending AWHB. 
 However, their suboptimal performance across all metrics suggests that it is inadequate for addressing the recommendation task. 
 Similar to the result in RQ1, the better performance of \textbf{Random Forest} also suggests the features extracted by UniXcoder are more suitable for actionable warning recommendation than the selected features used in \textbf{Golden Feature}.

\textbf{Our proposed model is effective for AWHB recommendation and outperforms all the baseline methods.} The nDCG@1 value exceeds 0.3, indicating that over three-tenths of the queries can accurately identify AWHB in the topmost position of the recommended warning list with a high degree of certainty. The MRR value is approximately 0.45, suggesting that on average, the first AWHB can be found at the third position in the recommended warning list. We attribute this performance to the following reasons: First, the weak label from our matching rules can be efficient in distinguishing the real bugs from actionable warnings. Second, the semantic features employed in {\sc \textbf{ACWRecommendor}} are suitable for AWHB recommendation. To be specific, the commit message matching rule utilizes text information of warnings, and the code change matching rule relies on detecting similar patterns in the code information of warnings.

\subsubsection{RQ3: Type-Wise Evaluation}

\begin{table}
  \begin{center}
    \caption{Type-Wise Recommendation Evaluation}
    \vspace{-10pt}
    \small
    \label{Table:Type-Wise Recommendation Effectiveness Evaluation}
    \begin{tabular}{|c|c|c|c|c|c|} 
    \hline
    Measure & nDCG@1 & nDCG@3 & nDCG@5 & MRR \\
    \hline
    \hline
    Uninitialized Variable & 0.380 & 0.492 & 0.531 & 0.511  \\
    \hline
    Null Dereference & 0.413 & 0.507 & 0.557 & 0.520 \\
    \hline
    Resource Leak & 0.485 & 0.534 & 0.591 & 0.566 \\
    \hline
    Dead Store & 0.263 & 0.395 & 0.419 & 0.410  \\
    \hline
    Buffer Overflow &  0.312 & 0.425 & 0.487 & 0.434   \\
    \hline
    \end{tabular}
  \end{center}
\end{table}

To gain a deeper understanding of how our model performs on different static analysis tools and different warning types, in this research question we conduct a type-wise evaluation. 
Specifically, we calculated {\sc \textbf{ACWRecommender}}'s performance regarding five warning types (i.e., \textbf{Uninitialized Variable}, \textbf{Null Dereference}, \textbf{Resource Leak}, \textbf{Dead Store} and \textbf{Buffer Overflow}) of two static analysis tools (i.e., \textit{Infer} and \textit{Flawfinder}). 
The evaluation results are presented in Table~\ref{Table:Type-Wise Recommendation Effectiveness Evaluation}. 
From these tables, we have the following observations: 

\textbf{Our approach can generalize well to different static analysis tools and different warning types.} 
To verify the generalizability of our approach, we adopt two unique static analysis tools, i.e., \textit{Infer} and \textit{Flawfinder}, to cover different types of warnings. 
It is clear that {\sc \textbf{ACWRecommender}} performs consistently well for most types of warnings, such as \textbf{Uninitialized Variable}, \textbf{Null Dereference} and \textbf{Resource Leak} warning reported by \textit{Infer} and \textbf{Buffer Overflow} warning reported by \textit{Flawfinder}. 
We attribute the generalization ability of our approach to the following reasons: (i) Our base model UnixCoder can automatically and effectively learn features from the warning code and warning report without any reworking. 
Different static analysis tools' output can be easily incorporated into the pipeline of our approach. 
(ii) Our weak supervised labeling method establishes the structural and semantic matching rule for AWHBs, these rules are applicable for general warning types. 

\subsection{In-the-wild Evaluation}
Our ultimate goal is to help developers find real bugs reported by SA tools by inspecting as few warnings as possible. 
To do this, we further conduct an in-the-wild evaluation to evaluate the practical value of our {\sc ACWRecommender} for checking real bugs in real-world GitHub repositories. 
We began by randomly selecting 10 C repositories hosted on GitHub that had more than 500 stars and were last updated within a week. 
We ensured that the selected repositories were not used in the previous training and testing set. 
Next, we ran \textit{Infer} on the latest revision of each repository and obtained the \textit{Infer} reported warnings for each repository. 
Then {\sc ACWRecommendor} reranks the \textit{Infer} warnings and generates a warning list for checking. 
After removing 4 projects that can not be compiled, we obtained 2,197 \textit{Infer} reported warnings from 6 selected repositories (namely \texttt{libevent}, \texttt{flac}, \texttt{vifm}, \texttt{xrdp}, \texttt{netcdf-c} and \texttt{radare2}).

\noindent\textbf{Developer's Feedback of Pull Request}
\label{Sec:Developer Feedback of Pull Request}
Theoretically, we can submit all of our top-ranked warnings to ask developers to confirm if these warnings are real bugs. 
However, the workload is too heavy if developers need to check every warning candidate. 
To reduce the burden of developers, we first checked Top30\% (743) reported
warnings and validated 23 warnings as potential bugs. 
Following that, we manually fixed each validated warning and submitted a warning fix PR (Pull Request) directly to its corresponding GitHub repository.  
Among the 23 submitted validated warnings, 21 of them have already been confirmed by the developers, 19 of them have been merged, and 2 of them have been approved by code reviewers. 
It is worth mentioning that \textbf{more than half of the confirmed bugs (12/21) are identified within the Top 10\% of the recommended warnings, which justifies the effectiveness of our tool for alleviating the effort for finding real bugs in a massive amount of warnings.}
 
In general, developers appreciated the contributions we made to their repositories. 
For example, after confirming our validated \textbf{Null Dereference} warning~\cite{vifm/pull/882}, one developer of \texttt{vifm} project responded \textit{``Thanks. Also realized another issue with make\_node\(\) while reviewing the changes.''} 
The \textit{Infer} reported 239 warnings for this revision, it is time-consuming and costly for developers to check these warnings one by one, after adopting our tool, the developer can easily find this bug by only checking 3 top warnings. 
Similarly, another developer from \texttt{libevent} confirmed our pull-request~\cite{libevent/pull/1429}
commented after our further explanation, and commented that \textit{``Yes, missed the err label, you are right.''}
The response shows the AWHB recommended by our tool can assist developers in finding bugs that they may have overlooked.

\noindent\textbf{Developer's Feedback of Submitted Issues.}
Because we lack the programming context and domain expertise with respect to each test project, our validated warnings are by no means the entire warnings that need to be fixed. 
In other words, there are still warnings that we are unsure about. 
To alleviate the bias of our manual validation, for each top-10 reported warning in the above GitHub repositories, we formed the warning as an issue report and submitted the issue to its GitHub repository. 
In particular, we submitted 50 warnings (excluding 10 warnings already confirmed through pull requests) to 6 GitHub repositories. 
10 issues have been confirmed by developers as real bugs. 
Finally, \textbf{among these top-60 warnings reported by {\sc \textbf{ACWRecommender}}, 27 warnings (including 10 warnings submitted by pull requests and 17 warnings submitted by issue reports) are identified as real bugs}, which further confirms the effectiveness of our tool for recommending valid actionable warnings. 

The developers of \texttt{radare2} validated and fixed all the committed top-5 warnings\cite{radare2/issues}. 
The developers of \texttt{xrdp} validated two \textbf{Dead Store} warnings\cite{xrdp/security}. 
Some of our recommended warnings are true defects but do not affect the usage or safety of software, therefore the developers are reluctant to act on these warnings. 
For example, even \texttt{xrdp} developers confirmed our reported \textbf{Dead Store} warnings, they consider these warnings \textit{"does not matter"} and replied \textit{``I can't see that any of these are of any concern from a security perspective.''} 
The developers' perspective of \textbf{Dead Store} warnings has led to a low-quality label for this warning type, resulting in real bugs being labeled as false alarms. 
It greatly hinders the learning performance of our tool on this warning type and further explains the relatively poor performance of our approach on \textbf{Dead Store} warnings. 
Some of our recommended warnings are considered by developers as false alarms by overestimating program behaviors. 
For example, for a \textbf{Null Dereference} warning reported by our tool in \texttt{xrdp}, the developer responded \textit{"Potentially a NULL pointer dereference but xrdp is full of instances of this idiom."}

%% file: discussion.tex
\subsection{\xzp{Generalizability of Heuristic Rules}}
\xzp{In this work, our weakly-supervised labeling strategy is based on heuristically defined semantic and structural matching rules. 
These heuristic rules are successfully applied to \textit{Infer} and \textit{Flawfinder} to verify their effectiveness. 
To demonstrate whether these rules are generalizable to other SA tools, we extend these summarized rules to an additional static analysis tool, \textit{SonarQube}.  
}

\noindent{\xzp{\textbf{Experimental Setup.}}}
\xzp{We randomly sampled 50 projects from our dataset, for each project, we ran \textit{SonarQube} on its compiled code of each commit version to collect actionable warnings (introduced in Section~\ref{sec:approach}). 
As a result, we collected 169 actionable warnings across the aforementioned five warning types. 
After that, we apply our heuristic rules to these actionable warnings to identify AWHBs. 
Finally, 56 out of 169 actionable warnings are labeled as AWHBs. 
After that, we follow the same experimental settings (detailed in Section~\ref{rq1}) to manually review each of the 169 actionable warnings as real bugs or not. 
}

\noindent{\xzp{\textbf{Experimental Results.}}}
\xzp{
The experimental results are shown in Table~\ref{Table:SonarQube Warnings}. 
From the table, we can see that: (1) 83.9\% AWHBs (i.e., 47 out of 56) were confirmed by developers as real bugs. 
The effectiveness of our heuristic rules is consistent across different static analysis tools (i.e., \textit{Infer}, \textit{Flawfinder}, and \textit{SonarQube}), demonstrating the effectiveness and robustness of our weak supervision labeling strategy. 
(2) It is worth mentioning that we use warning descriptions and ASTs to define these heuristic rules, without relying on any specific static analysis tools. 
This design ensures our heuristic rules are highly generalizable and scalable, which can be easily extended to other static analysis tools. 
}

\begin{table} 
\small
  \begin{center}
\xzp{
    \caption{Labeling Effectiveness on SonarQube Warnings}
    \vspace{-5pt}
    \label{Table:SonarQube Warnings}
    \begin{tabular}{|c|c|c|c|c|c|c|}
    \hline
    Count & \makecell[c]{Actionable\\Warnings} & AWHB & Real Bugs & Ratio \\
    \hline
    \hline
    Uninitialized Variable & 69 & 29 & 25 & 25\slash29 \\
    \hline
    Null Dereference & 47 & 19 & 15 & 15\slash19 \\
    \hline
    Resource Leak & 28 & 5 & 5 & 5\slash5 \\
    \hline
    Dead Store & 14 & 0  & 0 & - \\
    \hline
    Buffer Overflow & 11 & 3 & 2 & 2\slash3 \\
    \hline
    Total & 169 & 56 & 47 & 83.9\% \\
    \hline
    \end{tabular}
    }
  \end{center}
\end{table}

\subsection{\xzp{Practial Application Settings}}
\xzp{Our approach re-ranked AWHBs higher up for developers' investigation. To assess the trade-off between inspection effort and the tool's effectiveness, we analyze the practical application settings.} 
\noindent{\xzp{\textbf{Experimental Setup.}}}
\xzp{
To assess the trade-off between inspection effort and effectiveness of identifying real bugs, we evaluate the $Recall@K$ metric using our testing dataset, where $K$ refers to the top-K\% ranked warnings, and $Recall@K$ represents the proportion of real bugs that appear in top-K\% ranked warnings. 
}

\begin{figure}
	\centering
	\includegraphics[width=0.9\linewidth]{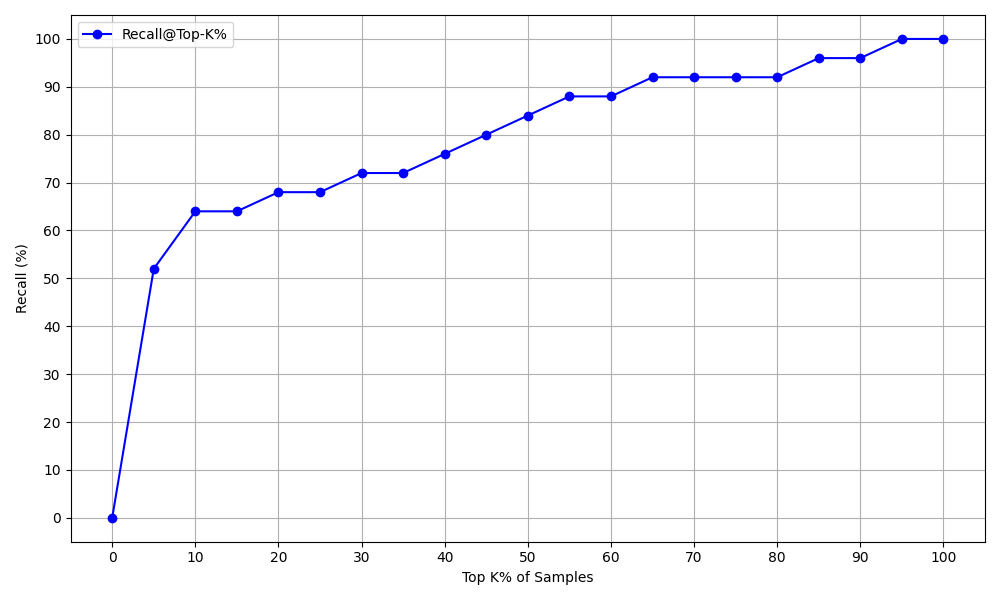}
	\caption{Recall@Top-K\% curve}
    \label{Fig:Recall@K curve}
\end{figure}

\noindent{\xzp{\textbf{Experimental Results.}}}
\xzp{
The evaluation result is shown in Fig.~\ref{Fig:Recall@K curve}. 
From the figure, we can see that: (1) Inspecting the top 5\% of ranked warnings can already cover over 50\% of AWHB. Therefore, in practical usage, we recommend that developers review the top 5\% of recommended warnings to achieve a good balance between inspection cost and bug detection effectiveness. 
(2) As shown in our in-the-wild experiments, we submitted the top 10 ranked warnings identified by our approach, and 27/60 submitted warnings were confirmed as real bugs by project developers, demonstrating strong real-world applicability. 
(3) For deployment, our approach can be easily implemented as a plugin or extension for popular IDEs (e.g., VSCode, IntelliJ). 
Once integrated, it can analyze static analysis warnings in real time and highlight only the AWHB. 
It can also be deployed as a post-processing step in CI pipelines to filter SA outputs before they reach developers. 
This helps developers better utilize SA tools and further improve software quality. 
}

\subsection{\xzp{Reducing Manual Efforts with LLMs}}
\xzp{
Our approach relies on heuristic rules in the weak supervision labeling process, while defining these rules involves manual inspection, we believe the effort is both cost-effective and efficient. 
For semantic rules, relevant keywords for each warning type can be easily extracted from warning descriptions, making this step straightforward and low-effort. 
For structural rules, although summarizing fix patterns from ASTs requires domain expertise, this manual work is a one-time cost. 
Once established, these patterns can be reused across projects with minimal maintenance. 
}

\noindent{\xzp{\textbf{Experimental Setup.}}}
\xzp{Inspired by the capabilities of LLMs~\cite{dai2025less, xue2024selfpico, mai2024human, dai2024mpcoder, yan2023closer, mai2025towards}, we further explored how LLMs can be utilized to reduce manual efforts during this process. 
Specifically, for each type of warning, we leverage ChatGPT to automatically generate relevant keywords (i.e., semantic rules) and candidate fix patterns (i.e., structural rules) by providing LLMs with reported warnings. 
}

\noindent{\xzp{\textbf{Experimental Results.}}}
\xzp{
The detailed prompts and LLMs' outputs are demonstrated in our recorded chat histories~\cite{null_gpt,dead_gpt,uninit_gpt,leak_gpt,overflow_gpt}. 
We manually reviewed the LLM outputs to assess their consistency with human-summarized patterns. 
Our manual analysis reveals that: (1) Regarding semantic rules, the keywords produced by the LLM cover most of the human summarized keywords across all five types of warnings. 
(2) Regarding structural rules, three types of warnings (e.g., Dead Code, Null Dereference, and Uninitialized Variable) exactly match our human-defined patterns and even suggest additional fix patterns. 
For Resource Leak and Buffer Overflow warnings, the LLM generates specific fix patterns tailored to typical cases.
For example, they recommend using \texttt{fclose()} for \texttt{\_IO\_FILE} leaks but may overlook user-defined resource-closing functions. 
In contrast, more comprehensive function call patterns, e.g., \textit{close} and \textit{free}, are summarized by human experts. 
(3) Overall, LLMs show promising potential in reducing manual efforts for summarizing heuristic rules, and how to utilize LLMs to generate comprehensive and validated rules remains an open challenge and is beyond the scope of this work.
}

\subsection{\xzp{Threats to Validity}}
\xzp{\textbf{Threats to internal validity.}}
\xzp{The internal validity relates to potential errors in model implementation and experimental settings. 
Our approach relies on heuristic rules for data labeling, which may introduce labeling noise. 
To mitigate this, we manually reviewed the automatically labeled data and found that over 80\% of the labels were accurate.
Our heuristic-based method also suffered from overlooking false negatives. 
Warnings present in the latest commit, but potentially fixed in the future, are not labeled as actionable under our current definition. 
These false negatives are excluded from our analysis because they have not yet met the actionable criterion. Additionally, some false negatives arise from the incompleteness of our manually summarized heuristic rules, which may fail to capture all possible fix patterns.
In the future, we plan to enhance our labeling strategy to better capture delayed fixes. 
}

\noindent\xzp{\textbf{Threats to external validity.} }
\xzp{External validity concerns the generalizability and representativeness of our results. 
Our experiments are conducted on five common C/C++ vulnerability types detected by two widely used static analysis tools. 
In terms of representativeness, our study focuses on five representative warning types. 
Four of them (Uninitialized Variable, Null Dereference, Resource Leak, Dead Store) account for over 99\% of Infer’s reported warnings, while Buffer Overflow is the most frequent warning in Flawfinder. 
Regarding severity, several selected types correspond to top-ranked CWEs~\cite{TOPCWE} and have been widely studied~\cite{malavolta2023javascript, tu2024beyond, li2023assisting,li2022path}. 
In terms of generalizability, our framework is designed to be both generalizable and scalable, and the consistent effectiveness across different SA tools (i.e., \textit{Infer}, \textit{Flawfinder}, and \textit{SonarQube}) further demonstrates the generalizability of our approach. 
}

\noindent\xzp{\noindent\textbf{Threats to construct validity.} }
\xzp{The Construct validity is introduced by the confirmation bias in manual evaluations, especially when assessing the accuracy of our weak supervision labeling strategy. To mitigate this risk, we recruited three experienced developers and presented the warnings in a randomized order, asking each to determine whether each warning corresponded to a real bug independently.
}

%% file: relatedwork.tex
Based on the assumption that the warnings that are not acted upon by developers are regarded as false alarms. The goal of warning identification is to identify actionable warnings and false alarms. Several studies, leveraged d machine learning to distinguish actionable and non-actionable alarms based on manually predefined feature~\cite{hanam2014finding, yuksel2013automated, Utture2022StrikingAB, Alikhashashneh2018UsingML, Heo2017MachineLearningGuidedSU, Koc2017LearningAC, Lee2019ClassifyingFP, Zhang2019AVA, Liang2010AutomaticCO}. 
Zhang et al.~\cite{Zhang2019AVA} extracted variable-level features and showed that they outperform file/class-level ones.
Wang et al.~\cite{wang2018there} identified 20 “Golden Features,” achieving state-of-the-art performance in actionable warning identification.
However, some researchers doubted the result of the Golden Feature. Yang et al.\cite{Yang2021LearningTR} investigated the superior efficacy of the Golden Feature in SVM, as compared to CNN, due to the intrinsically simple data. Kang et al.~\cite{kang2022detecting} revealed a data leakage in Golden Feature, and reported a pool performance for it after fixing the data leakage and data duplication. They concluded that the Golden Feature is inadequate for the warning identification task.

The research on warning recommendation aims to prioritize warnings that are more likely to be true bugs and ordered up in the list. The key insight of these studies is to predict the probability of static analysis warnings. 
Some of them combined the results from several SA tools~\cite{ribeiro2019ranking, Flynn2018PrioritizingAF, Nunes2018AnES, Meng2008AnAT}.
For example, Ribeiro et al.~\cite{ribeiro2019ranking} using AdaBoost on features from three tools, and Flynn et al.~\cite{Flynn2018PrioritizingAF} gathered warnings from a suite of software assurance tools.
However, they require analyzing the same code by multiple tools, and this increases the code analysis time.
Some researchers recommend the warnings involving additional human-in-the-loop~\cite{raghothaman2018user, Heo2019ContinuouslyRA, kim2022learning, Ngo2021RankingWO}.
Mangal et al. ~\cite{raghothaman2018user} used Bayesian inference on a probabilistic model derived from the derivation graph, and regarded the warning inspected by the developer as the ground truth.
{\sc Drake}~\cite{Heo2019ContinuouslyRA}, for instance, updated weights iteratively based on user feedback.
It updated the weights based on the user feedback each round. 
A number of studies recommend the warnings based on the history-aware information~\cite{Aman2019ASA, Burhandenny2017InvestigationOC, Kim2007PrioritizingWC, Williams2005AutomaticMO}.
Aman et al. ~\cite{Aman2019ASA} estimated the lifetimes of alarms by using the survival analysis method, and assigned higher priority to alarms that have shorter lifetimes.

All the prior approaches focus on handling actionable warnings.
While our approach first proposed a weakly supervised labeling method to assign different labels to the actionable warnings based on their probabilities to be real bugs. 
This research is based on our preliminary tool demo~\cite{xue2023acwrecommender} by extending it to 
a two-stage modeling, conducting a comprehensive evaluation to verify its effectiveness and value.  

%% file: conclusion.tex
This research aims to identify actionable warnings produced by static analysis tool and recommend the Actionable Warning with High probability to be real Bug. 
To address these task, we first collect actionable warnings from top-500 GitHub repositories. 
We then propose a weakly supervised strategy to identify \textbf{A}ctionable \textbf{W}arning with \textbf{H}igh probability to be real \textbf{B}ugs (\textbf{AWHBs}). 
We propose an approach, namely {\sc ACWRecommender}, which leverages UniXcoder to recommend AWHBs. 
Comprehensive evaluation results show the effectiveness of our approach on this task. 
Extensive experiments on the real-world GitHub repositories have demonstrated its practical value in improving the utility of SA tools. 

%% file: main.bbl

\begin{thebibliography}{79}


\ifx \showCODEN    \undefined \def \showCODEN     #1{\unskip}     \fi
\ifx \showISBNx    \undefined \def \showISBNx     #1{\unskip}     \fi
\ifx \showISBNxiii \undefined \def \showISBNxiii  #1{\unskip}     \fi
\ifx \showISSN     \undefined \def \showISSN      #1{\unskip}     \fi
\ifx \showLCCN     \undefined \def \showLCCN      #1{\unskip}     \fi
\ifx \shownote     \undefined \def \shownote      #1{#1}          \fi
\ifx \showarticletitle \undefined \def \showarticletitle #1{#1}   \fi
\ifx \showURL      \undefined \def \showURL       {\relax}        \fi
\providecommand\bibfield[2]{#2}
\providecommand\bibinfo[2]{#2}
\providecommand\natexlab[1]{#1}
\providecommand\showeprint[2][]{arXiv:#2}

\bibitem[vif(2023)]%
        {vifm/pull/882}
 \bibinfo{year}{2023}\natexlab{}.
\newblock
\urldef\tempurl%
\url{https://github.com/vifm/vifm/pull/882}
\showURL{%
\tempurl}


\bibitem[lib(2023a)]%
        {libevent/pull/1429}
 \bibinfo{year}{2023}\natexlab{a}.
\newblock
\urldef\tempurl%
\url{https://github.com/libevent/libevent/pull/1429}
\showURL{%
\tempurl}


\bibitem[rad(2023)]%
        {radare2/issues}
 \bibinfo{year}{2023}\natexlab{}.
\newblock
\urldef\tempurl%
\url{https://github.com/radareorg/radare2/issues/22313}
\showURL{%
\tempurl}


\bibitem[xrd(2023)]%
        {xrdp/security}
 \bibinfo{year}{2023}\natexlab{}.
\newblock
\urldef\tempurl%
\url{https://github.com/neutrinolabs/xrdp/security/advisories/GHSA-f3qh-qc3p-254c#event-113510}
\showURL{%
\tempurl}


\bibitem[lib(2023b)]%
        {libevhtp/commit/d13b72b}
 \bibinfo{year}{2023}\natexlab{b}.
\newblock \bibinfo{title}{Actionable warning in libevhtp}.
\newblock
\urldef\tempurl%
\url{https://github.com/Yellow-Camper/libevhtp/commit/d13b72b3887096f54d3570b723f2f518cf0e903b}
\showURL{%
\tempurl}


\bibitem[ope(2023)]%
        {open62541/commit/d52786e}
 \bibinfo{year}{2023}\natexlab{}.
\newblock \bibinfo{title}{Actionable warning in open62541}.
\newblock
\urldef\tempurl%
\url{https://github.com/open62541/open62541/commit/d52786e61878fd4b2b798c84de49c289b2bfff6f}
\showURL{%
\tempurl}


\bibitem[TOP(2024)]%
        {TOPCWE}
 \bibinfo{year}{2024}\natexlab{}.
\newblock \bibinfo{title}{2024 CWE Top 25 Most Dangerous Software Weaknesses}.
\newblock
\urldef\tempurl%
\url{https://cwe.mitre.org/top25/archive/2024/2024_cwe_top25.html}
\showURL{%
\tempurl}


\bibitem[rep(2025)]%
        {replication}
 \bibinfo{year}{2025}\natexlab{}.
\newblock \bibinfo{title}{ACWRecommender replication package}.
\newblock
\urldef\tempurl%
\url{https://zenodo.org/records/15550900}
\showURL{%
\tempurl}


\bibitem[ove(2025)]%
        {overflow_gpt}
 \bibinfo{year}{2025}\natexlab{}.
\newblock \bibinfo{title}{Buffer Overflow Pattern from ChatGPT}.
\newblock
\urldef\tempurl%
\url{https://chatgpt.com/share/686f813f-e080-800d-aee2-2b94db9589a7}
\showURL{%
\tempurl}


\bibitem[dea(2025)]%
        {dead_gpt}
 \bibinfo{year}{2025}\natexlab{}.
\newblock \bibinfo{title}{Dead Store Pattern from ChatGPT}.
\newblock
\urldef\tempurl%
\url{https://chatgpt.com/share/686f7f58-98a4-800d-92f5-a02f51edf534}
\showURL{%
\tempurl}


\bibitem[nul(2025)]%
        {null_gpt}
 \bibinfo{year}{2025}\natexlab{}.
\newblock \bibinfo{title}{Null Dereference Pattern from ChatGPT}.
\newblock
\urldef\tempurl%
\url{https://chatgpt.com/share/68387613-fdf8-800d-824f-92ea172ec4df}
\showURL{%
\tempurl}


\bibitem[lea(2025)]%
        {leak_gpt}
 \bibinfo{year}{2025}\natexlab{}.
\newblock \bibinfo{title}{Reasource Leak Pattern from ChatGPT}.
\newblock
\urldef\tempurl%
\url{https://chatgpt.com/share/686f8015-3acc-800d-b187-cbe7554cdc55}
\showURL{%
\tempurl}


\bibitem[uni(2025)]%
        {uninit_gpt}
 \bibinfo{year}{2025}\natexlab{}.
\newblock \bibinfo{title}{Uninitialized Variable Pattern from ChatGPT}.
\newblock
\urldef\tempurl%
\url{https://chatgpt.com/share/686f7e7c-006c-800d-89c2-2a60747cf4cf}
\showURL{%
\tempurl}


\bibitem[Alikhashashneh et~al\mbox{.}(2018)]%
        {Alikhashashneh2018UsingML}
\bibfield{author}{\bibinfo{person}{Enas~A. Alikhashashneh}, \bibinfo{person}{Rajeev~R. Raje}, {and} \bibinfo{person}{James~H. Hill}.} \bibinfo{year}{2018}\natexlab{}.
\newblock \showarticletitle{Using Machine Learning Techniques to Classify and Predict Static Code Analysis Tool Warnings}.
\newblock \bibinfo{journal}{\emph{2018 IEEE/ACS 15th International Conference on Computer Systems and Applications (AICCSA)}} (\bibinfo{year}{2018}), \bibinfo{pages}{1--8}.
\newblock


\bibitem[Aman et~al\mbox{.}(2019)]%
        {Aman2019ASA}
\bibfield{author}{\bibinfo{person}{Hirohisa Aman}, \bibinfo{person}{Sousuke Amasaki}, \bibinfo{person}{Tomoyuki Yokogawa}, {and} \bibinfo{person}{Minoru Kawahara}.} \bibinfo{year}{2019}\natexlab{}.
\newblock \showarticletitle{A Survival Analysis-Based Prioritization of Code Checker Warning: A Case Study Using PMD}. In \bibinfo{booktitle}{\emph{International Conference on Big Data, Cloud Computing, Data Science \& Engineering}}.
\newblock


\bibitem[Avgustinov et~al\mbox{.}(2015)]%
        {avgustinov2015tracking}
\bibfield{author}{\bibinfo{person}{Pavel Avgustinov}, \bibinfo{person}{Arthur~I Baars}, \bibinfo{person}{Anders~S Henriksen}, \bibinfo{person}{Greg Lavender}, \bibinfo{person}{Galen Menzel}, \bibinfo{person}{Oege De~Moor}, \bibinfo{person}{Max Schafer}, {and} \bibinfo{person}{Julian Tibble}.} \bibinfo{year}{2015}\natexlab{}.
\newblock \showarticletitle{Tracking static analysis violations over time to capture developer characteristics}. In \bibinfo{booktitle}{\emph{2015 IEEE/ACM 37th IEEE International Conference on Software Engineering}}, Vol.~\bibinfo{volume}{1}. IEEE, \bibinfo{pages}{437--447}.
\newblock


\bibitem[Brat and Venet(2005)]%
        {brat2005precise}
\bibfield{author}{\bibinfo{person}{Guillaume Brat} {and} \bibinfo{person}{Arnaud Venet}.} \bibinfo{year}{2005}\natexlab{}.
\newblock \showarticletitle{Precise and scalable static program analysis of NASA flight software}. In \bibinfo{booktitle}{\emph{2005 IEEE Aerospace Conference}}. IEEE, \bibinfo{pages}{1--10}.
\newblock


\bibitem[Burhandenny et~al\mbox{.}(2017)]%
        {Burhandenny2017InvestigationOC}
\bibfield{author}{\bibinfo{person}{Aji~Ery Burhandenny}, \bibinfo{person}{Hirohisa Aman}, {and} \bibinfo{person}{Minoru Kawahara}.} \bibinfo{year}{2017}\natexlab{}.
\newblock \showarticletitle{Investigation of Coding Violations Focusing on Authorships of Source Files}.
\newblock \bibinfo{journal}{\emph{2017 5th Intl Conf on Applied Computing and Information Technology/4th Intl Conf on Computational Science/Intelligence and Applied Informatics/2nd Intl Conf on Big Data, Cloud Computing, Data Science (ACIT-CSII-BCD)}} (\bibinfo{year}{2017}), \bibinfo{pages}{248--253}.
\newblock


\bibitem[Calcagno et~al\mbox{.}(2015)]%
        {Calcagno2015MovingFW}
\bibfield{author}{\bibinfo{person}{Cristiano Calcagno}, \bibinfo{person}{Dino Distefano}, \bibinfo{person}{J{\'e}r{\'e}my Dubreil}, \bibinfo{person}{Dominik Gabi}, \bibinfo{person}{Pieter Hooimeijer}, \bibinfo{person}{Martino Luca}, \bibinfo{person}{Peter~W. O'Hearn}, \bibinfo{person}{Irene Papakonstantinou}, \bibinfo{person}{Jim Purbrick}, {and} \bibinfo{person}{Dulma Rodriguez}.} \bibinfo{year}{2015}\natexlab{}.
\newblock \showarticletitle{Moving Fast with Software Verification}. In \bibinfo{booktitle}{\emph{NASA Formal Methods}}.
\newblock
\urldef\tempurl%
\url{https://api.semanticscholar.org/CorpusID:12281434}
\showURL{%
\tempurl}


\bibitem[Chen et~al\mbox{.}(2024)]%
        {10.1145/3597503.3639583}
\bibfield{author}{\bibinfo{person}{Zirui Chen}, \bibinfo{person}{Xing Hu}, \bibinfo{person}{Xin Xia}, \bibinfo{person}{Yi Gao}, \bibinfo{person}{Tongtong Xu}, \bibinfo{person}{David Lo}, {and} \bibinfo{person}{Xiaohu Yang}.} \bibinfo{year}{2024}\natexlab{}.
\newblock \showarticletitle{Exploiting Library Vulnerability via Migration Based Automating Test Generation}. In \bibinfo{booktitle}{\emph{Proceedings of the IEEE/ACM 46th International Conference on Software Engineering}} (Lisbon, Portugal) \emph{(\bibinfo{series}{ICSE '24})}. \bibinfo{publisher}{Association for Computing Machinery}, \bibinfo{address}{New York, NY, USA}, Article \bibinfo{articleno}{228}, \bibinfo{numpages}{12}~pages.
\newblock
\showISBNx{9798400702174}
\href{https://doi.org/10.1145/3597503.3639583}{doi:\nolinkurl{10.1145/3597503.3639583}}


\bibitem[Dai et~al\mbox{.}(2025)]%
        {dai2025less}
\bibfield{author}{\bibinfo{person}{Zhenlong Dai}, \bibinfo{person}{Bingrui Chen}, \bibinfo{person}{Zhuoluo Zhao}, \bibinfo{person}{Xiu Tang}, \bibinfo{person}{Sai Wu}, \bibinfo{person}{Chang Yao}, \bibinfo{person}{Zhipeng Gao}, {and} \bibinfo{person}{Jingyuan Chen}.} \bibinfo{year}{2025}\natexlab{}.
\newblock \showarticletitle{Less is More: Adaptive Program Repair with Bug Localization and Preference Learning}. In \bibinfo{booktitle}{\emph{Proceedings of the AAAI Conference on Artificial Intelligence}}, Vol.~\bibinfo{volume}{39}. \bibinfo{pages}{128--136}.
\newblock


\bibitem[Dai et~al\mbox{.}(2024)]%
        {dai2024mpcoder}
\bibfield{author}{\bibinfo{person}{Zhenlong Dai}, \bibinfo{person}{Chang Yao}, \bibinfo{person}{WenKang Han}, \bibinfo{person}{Yuanying Yuanying}, \bibinfo{person}{Zhipeng Gao}, {and} \bibinfo{person}{Jingyuan Chen}.} \bibinfo{year}{2024}\natexlab{}.
\newblock \showarticletitle{Mpcoder: Multi-user personalized code generator with explicit and implicit style representation learning}. In \bibinfo{booktitle}{\emph{Proceedings of the 62nd Annual Meeting of the Association for Computational Linguistics (Volume 1: Long Papers)}}. \bibinfo{pages}{3765--3780}.
\newblock


\bibitem[Distefano et~al\mbox{.}(2006)]%
        {Distefano2006ALS}
\bibfield{author}{\bibinfo{person}{Dino Distefano}, \bibinfo{person}{Peter~W. O'Hearn}, {and} \bibinfo{person}{Hongseok Yang}.} \bibinfo{year}{2006}\natexlab{}.
\newblock \showarticletitle{A Local Shape Analysis Based on Separation Logic}. In \bibinfo{booktitle}{\emph{International Conference on Tools and Algorithms for Construction and Analysis of Systems}}.
\newblock
\urldef\tempurl%
\url{https://api.semanticscholar.org/CorpusID:361092}
\showURL{%
\tempurl}


\bibitem[Flynn et~al\mbox{.}(2018)]%
        {Flynn2018PrioritizingAF}
\bibfield{author}{\bibinfo{person}{Lori Flynn}, \bibinfo{person}{William Snavely}, \bibinfo{person}{David Svoboda}, \bibinfo{person}{Nathan~M. VanHoudnos}, \bibinfo{person}{Richard Qin}, \bibinfo{person}{Jennifer Burns}, \bibinfo{person}{David Zubrow}, \bibinfo{person}{Robert Stoddard}, {and} \bibinfo{person}{Guillermo Marce-Santurio}.} \bibinfo{year}{2018}\natexlab{}.
\newblock \showarticletitle{Prioritizing Alerts from Multiple Static Analysis Tools, Using Classification Models}.
\newblock \bibinfo{journal}{\emph{2018 IEEE/ACM 1st International Workshop on Software Qualities and their Dependencies (SQUADE)}} (\bibinfo{year}{2018}), \bibinfo{pages}{13--20}.
\newblock


\bibitem[Gao et~al\mbox{.}(2020)]%
        {gao2020technical}
\bibfield{author}{\bibinfo{person}{Zhipeng Gao}, \bibinfo{person}{Xin Xia}, \bibinfo{person}{David Lo}, {and} \bibinfo{person}{John Grundy}.} \bibinfo{year}{2020}\natexlab{}.
\newblock \showarticletitle{Technical Q8A site answer recommendation via question boosting}.
\newblock \bibinfo{journal}{\emph{ACM Transactions on Software Engineering and Methodology (TOSEM)}} \bibinfo{volume}{30}, \bibinfo{number}{1} (\bibinfo{year}{2020}), \bibinfo{pages}{1--34}.
\newblock


\bibitem[Gao et~al\mbox{.}(2021)]%
        {gao2021automating}
\bibfield{author}{\bibinfo{person}{Zhipeng Gao}, \bibinfo{person}{Xin Xia}, \bibinfo{person}{David Lo}, \bibinfo{person}{John Grundy}, {and} \bibinfo{person}{Thomas Zimmermann}.} \bibinfo{year}{2021}\natexlab{}.
\newblock \showarticletitle{Automating the removal of obsolete TODO comments}. In \bibinfo{booktitle}{\emph{Proceedings of the 29th ACM Joint Meeting on European Software Engineering Conference and Symposium on the Foundations of Software Engineering}}. \bibinfo{pages}{218--229}.
\newblock


\bibitem[Gao et~al\mbox{.}(2022)]%
        {Gao2022IKW}
\bibfield{author}{\bibinfo{person}{Zhipeng Gao}, \bibinfo{person}{Xin Xia}, \bibinfo{person}{D. Lo}, \bibinfo{person}{John~C. Grundy}, \bibinfo{person}{Xindong Zhang}, {and} \bibinfo{person}{Zhenchang Xing}.} \bibinfo{year}{2022}\natexlab{}.
\newblock \showarticletitle{I Know What You Are Searching For: Code Snippet Recommendation from Stack Overflow Posts}.
\newblock \bibinfo{journal}{\emph{ACM Transactions on Software Engineering and Methodology}} (\bibinfo{year}{2022}).
\newblock


\bibitem[Gao et~al\mbox{.}(2024)]%
        {gao2024easy}
\bibfield{author}{\bibinfo{person}{Zhipeng Gao}, \bibinfo{person}{Zhipeng Xue}, \bibinfo{person}{Xing Hu}, \bibinfo{person}{Weiyi Shang}, {and} \bibinfo{person}{Xin Xia}.} \bibinfo{year}{2024}\natexlab{}.
\newblock \showarticletitle{Easy over Hard: A Simple Baseline for Test Failures Causes Prediction}. In \bibinfo{booktitle}{\emph{Companion Proceedings of the 32nd ACM International Conference on the Foundations of Software Engineering}}. \bibinfo{pages}{306--317}.
\newblock


\bibitem[Hanam et~al\mbox{.}(2014)]%
        {hanam2014finding}
\bibfield{author}{\bibinfo{person}{Quinn Hanam}, \bibinfo{person}{Lin Tan}, \bibinfo{person}{Reid Holmes}, {and} \bibinfo{person}{Patrick Lam}.} \bibinfo{year}{2014}\natexlab{}.
\newblock \showarticletitle{Finding patterns in static analysis alerts: improving actionable alert ranking}. In \bibinfo{booktitle}{\emph{Proceedings of the 11th working conference on mining software repositories}}. \bibinfo{pages}{152--161}.
\newblock


\bibitem[Heckman and Williams(2008)]%
        {heckman2008establishing}
\bibfield{author}{\bibinfo{person}{Sarah Heckman} {and} \bibinfo{person}{Laurie Williams}.} \bibinfo{year}{2008}\natexlab{}.
\newblock \showarticletitle{On establishing a benchmark for evaluating static analysis alert prioritization and classification techniques}. In \bibinfo{booktitle}{\emph{Proceedings of the Second ACM-IEEE international symposium on Empirical software engineering and measurement}}. \bibinfo{pages}{41--50}.
\newblock


\bibitem[Heckman and Williams(2009)]%
        {heckman2009model}
\bibfield{author}{\bibinfo{person}{Sarah Heckman} {and} \bibinfo{person}{Laurie Williams}.} \bibinfo{year}{2009}\natexlab{}.
\newblock \showarticletitle{A model building process for identifying actionable static analysis alerts}. In \bibinfo{booktitle}{\emph{2009 International conference on software testing verification and validation}}. IEEE, \bibinfo{pages}{161--170}.
\newblock


\bibitem[Heckman and Williams(2011)]%
        {HECKMAN2011363}
\bibfield{author}{\bibinfo{person}{Sarah Heckman} {and} \bibinfo{person}{Laurie Williams}.} \bibinfo{year}{2011}\natexlab{}.
\newblock \showarticletitle{A systematic literature review of actionable alert identification techniques for automated static code analysis}.
\newblock \bibinfo{journal}{\emph{Information and Software Technology}} \bibinfo{volume}{53}, \bibinfo{number}{4} (\bibinfo{year}{2011}), \bibinfo{pages}{363--387}.
\newblock
\showISSN{0950-5849}
\href{https://doi.org/10.1016/j.infsof.2010.12.007}{doi:\nolinkurl{10.1016/j.infsof.2010.12.007}}
\newblock
\shownote{Special section: Software Engineering track of the 24th Annual Symposium on Applied Computing}.


\bibitem[Heo et~al\mbox{.}(2017)]%
        {Heo2017MachineLearningGuidedSU}
\bibfield{author}{\bibinfo{person}{Kihong Heo}, \bibinfo{person}{Hakjoo Oh}, {and} \bibinfo{person}{Kwangkeun Yi}.} \bibinfo{year}{2017}\natexlab{}.
\newblock \showarticletitle{Machine-Learning-Guided Selectively Unsound Static Analysis}.
\newblock \bibinfo{journal}{\emph{2017 IEEE/ACM 39th International Conference on Software Engineering (ICSE)}} (\bibinfo{year}{2017}), \bibinfo{pages}{519--529}.
\newblock


\bibitem[Heo et~al\mbox{.}(2019)]%
        {Heo2019ContinuouslyRA}
\bibfield{author}{\bibinfo{person}{Kihong Heo}, \bibinfo{person}{Mukund Raghothaman}, \bibinfo{person}{Xujie Si}, {and} \bibinfo{person}{M. Naik}.} \bibinfo{year}{2019}\natexlab{}.
\newblock \showarticletitle{Continuously reasoning about programs using differential Bayesian inference}.
\newblock \bibinfo{journal}{\emph{Proceedings of the 40th ACM SIGPLAN Conference on Programming Language Design and Implementation}} (\bibinfo{year}{2019}).
\newblock


\bibitem[Imtiaz et~al\mbox{.}(2019)]%
        {Imtiaz2019ChallengesWR}
\bibfield{author}{\bibinfo{person}{Nasif Imtiaz}, \bibinfo{person}{Akond Ashfaque~Ur Rahman}, \bibinfo{person}{Effat Farhana}, {and} \bibinfo{person}{Laurie~A. Williams}.} \bibinfo{year}{2019}\natexlab{}.
\newblock \showarticletitle{Challenges with Responding to Static Analysis Tool Alerts}.
\newblock \bibinfo{journal}{\emph{2019 IEEE/ACM 16th International Conference on Mining Software Repositories (MSR)}} (\bibinfo{year}{2019}), \bibinfo{pages}{245--249}.
\newblock
\urldef\tempurl%
\url{https://api.semanticscholar.org/CorpusID:195298514}
\showURL{%
\tempurl}


\bibitem[Kang et~al\mbox{.}(2022)]%
        {kang2022detecting}
\bibfield{author}{\bibinfo{person}{Hong~Jin Kang}, \bibinfo{person}{Khai~Loong Aw}, {and} \bibinfo{person}{David Lo}.} \bibinfo{year}{2022}\natexlab{}.
\newblock \showarticletitle{Detecting false alarms from automatic static analysis tools: how far are we?}. In \bibinfo{booktitle}{\emph{Proceedings of the 44th International Conference on Software Engineering}}. \bibinfo{pages}{698--709}.
\newblock


\bibitem[Kim et~al\mbox{.}(2022)]%
        {kim2022learning}
\bibfield{author}{\bibinfo{person}{Hyunsu Kim}, \bibinfo{person}{Mukund Raghothaman}, {and} \bibinfo{person}{Kihong Heo}.} \bibinfo{year}{2022}\natexlab{}.
\newblock \showarticletitle{Learning probabilistic models for static analysis alarms}. In \bibinfo{booktitle}{\emph{Proceedings of the 44th International Conference on Software Engineering}}. \bibinfo{pages}{1282--1293}.
\newblock


\bibitem[Kim and Ernst(2007)]%
        {Kim2007PrioritizingWC}
\bibfield{author}{\bibinfo{person}{Sunghun Kim} {and} \bibinfo{person}{Michael~D. Ernst}.} \bibinfo{year}{2007}\natexlab{}.
\newblock \showarticletitle{Prioritizing Warning Categories by Analyzing Software History}.
\newblock \bibinfo{journal}{\emph{Fourth International Workshop on Mining Software Repositories (MSR'07:ICSE Workshops 2007)}} (\bibinfo{year}{2007}), \bibinfo{pages}{27--27}.
\newblock


\bibitem[Koc et~al\mbox{.}(2017)]%
        {Koc2017LearningAC}
\bibfield{author}{\bibinfo{person}{Ugur Koc}, \bibinfo{person}{Parsa Saadatpanah}, \bibinfo{person}{Jeffrey~S. Foster}, {and} \bibinfo{person}{Adam~A. Porter}.} \bibinfo{year}{2017}\natexlab{}.
\newblock \showarticletitle{Learning a classifier for false positive error reports emitted by static code analysis tools}.
\newblock \bibinfo{journal}{\emph{Proceedings of the 1st ACM SIGPLAN International Workshop on Machine Learning and Programming Languages}} (\bibinfo{year}{2017}).
\newblock


\bibitem[Lee et~al\mbox{.}(2019)]%
        {Lee2019ClassifyingFP}
\bibfield{author}{\bibinfo{person}{Seongmin Lee}, \bibinfo{person}{Shin Hong}, \bibinfo{person}{Jungbae Yi}, \bibinfo{person}{Taeksu Kim}, \bibinfo{person}{Chul-Joo Kim}, {and} \bibinfo{person}{Shin Yoo}.} \bibinfo{year}{2019}\natexlab{}.
\newblock \showarticletitle{Classifying False Positive Static Checker Alarms in Continuous Integration Using Convolutional Neural Networks}.
\newblock \bibinfo{journal}{\emph{2019 12th IEEE Conference on Software Testing, Validation and Verification (ICST)}} (\bibinfo{year}{2019}), \bibinfo{pages}{391--401}.
\newblock


\bibitem[Li et~al\mbox{.}(2023)]%
        {li2023assisting}
\bibfield{author}{\bibinfo{person}{Haonan Li}, \bibinfo{person}{Yu Hao}, \bibinfo{person}{Yizhuo Zhai}, {and} \bibinfo{person}{Zhiyun Qian}.} \bibinfo{year}{2023}\natexlab{}.
\newblock \showarticletitle{Assisting static analysis with large language models: A chatgpt experiment}. In \bibinfo{booktitle}{\emph{Proceedings of the 31st ACM Joint European Software Engineering Conference and Symposium on the Foundations of Software Engineering}}. \bibinfo{pages}{2107--2111}.
\newblock


\bibitem[Li and Yang(2024)]%
        {li2024tracking}
\bibfield{author}{\bibinfo{person}{Junjie Li} {and} \bibinfo{person}{Jinqiu Yang}.} \bibinfo{year}{2024}\natexlab{}.
\newblock \showarticletitle{Tracking the Evolution of Static Code Warnings: The State-of-the-Art and a Better Approach}.
\newblock \bibinfo{journal}{\emph{IEEE Transactions on Software Engineering}} \bibinfo{volume}{50}, \bibinfo{number}{3} (\bibinfo{year}{2024}), \bibinfo{pages}{534--550}.
\newblock


\bibitem[Li et~al\mbox{.}(2022)]%
        {li2022path}
\bibfield{author}{\bibinfo{person}{Tuo Li}, \bibinfo{person}{Jia-Ju Bai}, \bibinfo{person}{Yulei Sui}, {and} \bibinfo{person}{Shi-Min Hu}.} \bibinfo{year}{2022}\natexlab{}.
\newblock \showarticletitle{Path-sensitive and alias-aware typestate analysis for detecting OS bugs}. In \bibinfo{booktitle}{\emph{Proceedings of the 27th ACM International Conference on Architectural Support for Programming Languages and Operating Systems}}. \bibinfo{pages}{859--872}.
\newblock


\bibitem[Liang et~al\mbox{.}(2010)]%
        {Liang2010AutomaticCO}
\bibfield{author}{\bibinfo{person}{Guangtai Liang}, \bibinfo{person}{Lingjing Wu}, \bibinfo{person}{Qian Wu}, \bibinfo{person}{Qianxiang Wang}, \bibinfo{person}{Tao Xie}, {and} \bibinfo{person}{Hong Mei}.} \bibinfo{year}{2010}\natexlab{}.
\newblock \showarticletitle{Automatic construction of an effective training set for prioritizing static analysis warnings}.
\newblock \bibinfo{journal}{\emph{Proceedings of the 25th IEEE/ACM International Conference on Automated Software Engineering}} (\bibinfo{year}{2010}).
\newblock


\bibitem[Liu et~al\mbox{.}(2021)]%
        {liu2021mining}
\bibfield{author}{\bibinfo{person}{Kui Liu}, \bibinfo{person}{Dongsun Kim}, \bibinfo{person}{Tegawend{\'e}~F. Bissyand{\'e}}, \bibinfo{person}{Shin Yoo}, {and} \bibinfo{person}{Yves Le~Traon}.} \bibinfo{year}{2021}\natexlab{}.
\newblock \showarticletitle{Mining Fix Patterns for FindBugs Violations}.
\newblock \bibinfo{journal}{\emph{IEEE Transactions on Software Engineering}} \bibinfo{volume}{47}, \bibinfo{number}{1} (\bibinfo{year}{2021}), \bibinfo{pages}{165--188}.
\newblock
\urldef\tempurl%
\url{https://doi.org/10.1109/TSE.2018.2884955}
\showURL{%
\tempurl}


\bibitem[Liu et~al\mbox{.}(2019)]%
        {liu2019avatar}
\bibfield{author}{\bibinfo{person}{Kui Liu}, \bibinfo{person}{Anil Koyuncu}, \bibinfo{person}{Dongsun Kim}, {and} \bibinfo{person}{Tegawend{\'e}~F Bissyand{\'e}}.} \bibinfo{year}{2019}\natexlab{}.
\newblock \showarticletitle{Avatar: Fixing semantic bugs with fix patterns of static analysis violations}. In \bibinfo{booktitle}{\emph{2019 IEEE 26th International Conference on Software Analysis, Evolution and Reengineering (SANER)}}. IEEE, \bibinfo{pages}{1--12}.
\newblock


\bibitem[Liu et~al\mbox{.}(2018)]%
        {liu2018neural}
\bibfield{author}{\bibinfo{person}{Zhongxin Liu}, \bibinfo{person}{Xin Xia}, \bibinfo{person}{Ahmed~E Hassan}, \bibinfo{person}{David Lo}, \bibinfo{person}{Zhenchang Xing}, {and} \bibinfo{person}{Xinyu Wang}.} \bibinfo{year}{2018}\natexlab{}.
\newblock \showarticletitle{Neural-machine-translation-based commit message generation: how far are we?}. In \bibinfo{booktitle}{\emph{Proceedings of the 33rd ACM/IEEE international conference on automated software engineering}}. \bibinfo{pages}{373--384}.
\newblock


\bibitem[Ma et~al\mbox{.}(2025a)]%
        {ma2025swe}
\bibfield{author}{\bibinfo{person}{Yingwei Ma}, \bibinfo{person}{Rongyu Cao}, \bibinfo{person}{Yongchang Cao}, \bibinfo{person}{Yue Zhang}, \bibinfo{person}{Jue Chen}, \bibinfo{person}{Yibo Liu}, \bibinfo{person}{Yuchen Liu}, \bibinfo{person}{Binhua Li}, \bibinfo{person}{Fei Huang}, {and} \bibinfo{person}{Yongbin Li}.} \bibinfo{year}{2025}\natexlab{a}.
\newblock \showarticletitle{SWE-GPT: A Process-Centric Language Model for Automated Software Improvement}.
\newblock \bibinfo{journal}{\emph{Proceedings of the ACM on Software Engineering}} \bibinfo{volume}{2}, \bibinfo{number}{ISSTA} (\bibinfo{year}{2025}), \bibinfo{pages}{2362--2383}.
\newblock


\bibitem[Ma et~al\mbox{.}(2025b)]%
        {ma2025alibaba}
\bibfield{author}{\bibinfo{person}{Yingwei Ma}, \bibinfo{person}{Qingping Yang}, \bibinfo{person}{Rongyu Cao}, \bibinfo{person}{Binhua Li}, \bibinfo{person}{Fei Huang}, {and} \bibinfo{person}{Yongbin Li}.} \bibinfo{year}{2025}\natexlab{b}.
\newblock \showarticletitle{Alibaba lingmaagent: Improving automated issue resolution via comprehensive repository exploration}. In \bibinfo{booktitle}{\emph{Proceedings of the 33rd ACM International Conference on the Foundations of Software Engineering}}. \bibinfo{pages}{238--249}.
\newblock


\bibitem[Mai et~al\mbox{.}(2024)]%
        {mai2024human}
\bibfield{author}{\bibinfo{person}{Yubo Mai}, \bibinfo{person}{Zhipeng Gao}, \bibinfo{person}{Xing Hu}, \bibinfo{person}{Lingfeng Bao}, \bibinfo{person}{Yu Liu}, {and} \bibinfo{person}{JianLing Sun}.} \bibinfo{year}{2024}\natexlab{}.
\newblock \showarticletitle{Are human rules necessary? generating reusable apis with cot reasoning and in-context learning}.
\newblock \bibinfo{journal}{\emph{Proceedings of the ACM on Software Engineering}} \bibinfo{volume}{1}, \bibinfo{number}{FSE} (\bibinfo{year}{2024}), \bibinfo{pages}{2355--2377}.
\newblock


\bibitem[Mai et~al\mbox{.}(2025)]%
        {mai2025towards}
\bibfield{author}{\bibinfo{person}{Yubo Mai}, \bibinfo{person}{Zhipeng Gao}, \bibinfo{person}{Haoye Wang}, \bibinfo{person}{Tingting Bi}, \bibinfo{person}{Xing Hu}, \bibinfo{person}{Xin Xia}, {and} \bibinfo{person}{Jianling Sun}.} \bibinfo{year}{2025}\natexlab{}.
\newblock \showarticletitle{Towards Better Answers: Automated Stack Overflow Post Updating}. In \bibinfo{booktitle}{\emph{2025 IEEE/ACM 47th International Conference on Software Engineering (ICSE)}}. IEEE, \bibinfo{pages}{591--603}.
\newblock


\bibitem[Malavolta et~al\mbox{.}(2023)]%
        {malavolta2023javascript}
\bibfield{author}{\bibinfo{person}{Ivano Malavolta}, \bibinfo{person}{Kishan Nirghin}, \bibinfo{person}{Gian~Luca Scoccia}, \bibinfo{person}{Simone Romano}, \bibinfo{person}{Salvatore Lombardi}, \bibinfo{person}{Giuseppe Scanniello}, {and} \bibinfo{person}{Patricia Lago}.} \bibinfo{year}{2023}\natexlab{}.
\newblock \showarticletitle{Javascript dead code identification, elimination, and empirical assessment}.
\newblock \bibinfo{journal}{\emph{IEEE Transactions on Software Engineering}} \bibinfo{volume}{49}, \bibinfo{number}{7} (\bibinfo{year}{2023}), \bibinfo{pages}{3692--3714}.
\newblock


\bibitem[Meng et~al\mbox{.}(2008)]%
        {Meng2008AnAT}
\bibfield{author}{\bibinfo{person}{Na Meng}, \bibinfo{person}{Qianxiang Wang}, \bibinfo{person}{Qian Wu}, {and} \bibinfo{person}{Hong Mei}.} \bibinfo{year}{2008}\natexlab{}.
\newblock \showarticletitle{An Approach to Merge Results of Multiple Static Analysis Tools (Short Paper)}.
\newblock \bibinfo{journal}{\emph{2008 The Eighth International Conference on Quality Software}} (\bibinfo{year}{2008}), \bibinfo{pages}{169--174}.
\newblock


\bibitem[Ngo et~al\mbox{.}(2021)]%
        {Ngo2021RankingWO}
\bibfield{author}{\bibinfo{person}{Kien-Tuan Ngo}, \bibinfo{person}{Dinh-Truong Do}, \bibinfo{person}{Thu-Trang Nguyen}, {and} \bibinfo{person}{Hieu~Dinh Vo}.} \bibinfo{year}{2021}\natexlab{}.
\newblock \showarticletitle{Ranking Warnings of Static Analysis Tools Using Representation Learning}.
\newblock \bibinfo{journal}{\emph{2021 28th Asia-Pacific Software Engineering Conference (APSEC)}} (\bibinfo{year}{2021}), \bibinfo{pages}{327--337}.
\newblock


\bibitem[Nunes et~al\mbox{.}(2018)]%
        {Nunes2018AnES}
\bibfield{author}{\bibinfo{person}{Paulo Jorge~Costa Nunes}, \bibinfo{person}{Ib{\'e}ria Medeiros}, \bibinfo{person}{Jos{\'e} Fonseca}, \bibinfo{person}{Nuno~Ferreira Neves}, \bibinfo{person}{Miguel~Pupo Correia}, {and} \bibinfo{person}{Marco Paulo~Amorim Vieira}.} \bibinfo{year}{2018}\natexlab{}.
\newblock \showarticletitle{An empirical study on combining diverse static analysis tools for web security vulnerabilities based on development scenarios}.
\newblock \bibinfo{journal}{\emph{Computing}}  \bibinfo{volume}{101} (\bibinfo{year}{2018}), \bibinfo{pages}{161--185}.
\newblock


\bibitem[Qiu et~al\mbox{.}(2021)]%
        {qiu2021deep}
\bibfield{author}{\bibinfo{person}{Fangcheng Qiu}, \bibinfo{person}{Zhipeng Gao}, \bibinfo{person}{Xin Xia}, \bibinfo{person}{David Lo}, \bibinfo{person}{John Grundy}, {and} \bibinfo{person}{Xinyu Wang}.} \bibinfo{year}{2021}\natexlab{}.
\newblock \showarticletitle{Deep just-in-time defect localization}.
\newblock \bibinfo{journal}{\emph{IEEE Transactions on Software Engineering}} \bibinfo{volume}{48}, \bibinfo{number}{12} (\bibinfo{year}{2021}), \bibinfo{pages}{5068--5086}.
\newblock


\bibitem[Raghothaman et~al\mbox{.}(2018)]%
        {raghothaman2018user}
\bibfield{author}{\bibinfo{person}{Mukund Raghothaman}, \bibinfo{person}{Sulekha Kulkarni}, \bibinfo{person}{Kihong Heo}, {and} \bibinfo{person}{Mayur Naik}.} \bibinfo{year}{2018}\natexlab{}.
\newblock \showarticletitle{User-guided program reasoning using Bayesian inference}. In \bibinfo{booktitle}{\emph{Proceedings of the 39th ACM SIGPLAN Conference on Programming Language Design and Implementation}}. \bibinfo{pages}{722--735}.
\newblock


\bibitem[Ribeiro et~al\mbox{.}(2019)]%
        {ribeiro2019ranking}
\bibfield{author}{\bibinfo{person}{Athos Ribeiro}, \bibinfo{person}{Paulo Meirelles}, \bibinfo{person}{Nelson Lago}, {and} \bibinfo{person}{Fabio Kon}.} \bibinfo{year}{2019}\natexlab{}.
\newblock \showarticletitle{Ranking warnings from multiple source code static analyzers via ensemble learning}. In \bibinfo{booktitle}{\emph{Proceedings of the 15th International Symposium on Open Collaboration}}. \bibinfo{pages}{1--10}.
\newblock


\bibitem[Smith et~al\mbox{.}(2019)]%
        {Smith2019HowDD}
\bibfield{author}{\bibinfo{person}{Justin Smith}, \bibinfo{person}{Brittany Johnson}, \bibinfo{person}{Emerson~R. Murphy-Hill}, \bibinfo{person}{Bill Chu}, {and} \bibinfo{person}{Heather~Richter Lipford}.} \bibinfo{year}{2019}\natexlab{}.
\newblock \showarticletitle{How Developers Diagnose Potential Security Vulnerabilities with a Static Analysis Tool}.
\newblock \bibinfo{journal}{\emph{IEEE Transactions on Software Engineering}}  \bibinfo{volume}{45} (\bibinfo{year}{2019}), \bibinfo{pages}{877--897}.
\newblock
\urldef\tempurl%
\url{https://api.semanticscholar.org/CorpusID:64711495}
\showURL{%
\tempurl}


\bibitem[Song et~al\mbox{.}(2022)]%
        {Song2022AdaptiveRS}
\bibfield{author}{\bibinfo{person}{Linxin Song}, \bibinfo{person}{Jieyu Zhang}, \bibinfo{person}{Tianxiang Yang}, {and} \bibinfo{person}{Masayuki Goto}.} \bibinfo{year}{2022}\natexlab{}.
\newblock \showarticletitle{Adaptive Ranking-based Sample Selection for Weakly Supervised Class-imbalanced Text Classification}. In \bibinfo{booktitle}{\emph{Conference on Empirical Methods in Natural Language Processing}}.
\newblock


\bibitem[Su et~al\mbox{.}(2021)]%
        {su2021reducing}
\bibfield{author}{\bibinfo{person}{Yanqi Su}, \bibinfo{person}{Zhenchang Xing}, \bibinfo{person}{Xin Peng}, \bibinfo{person}{Xin Xia}, \bibinfo{person}{Chong Wang}, \bibinfo{person}{Xiwei Xu}, {and} \bibinfo{person}{Liming Zhu}.} \bibinfo{year}{2021}\natexlab{}.
\newblock \showarticletitle{Reducing bug triaging confusion by learning from mistakes with a bug tossing knowledge graph}. In \bibinfo{booktitle}{\emph{2021 36th IEEE/ACM International Conference on Automated Software Engineering (ASE)}}. IEEE, \bibinfo{pages}{191--202}.
\newblock


\bibitem[Tahaei et~al\mbox{.}(2021)]%
        {Tahaei2021SecurityNI}
\bibfield{author}{\bibinfo{person}{Mohammad Tahaei}, \bibinfo{person}{Kami Vaniea}, \bibinfo{person}{Konstantin Beznosov}, {and} \bibinfo{person}{Maria~Klara Wolters}.} \bibinfo{year}{2021}\natexlab{}.
\newblock \showarticletitle{Security Notifications in Static Analysis Tools: Developers’ Attitudes, Comprehension, and Ability to Act on Them}.
\newblock \bibinfo{journal}{\emph{Proceedings of the 2021 CHI Conference on Human Factors in Computing Systems}} (\bibinfo{year}{2021}).
\newblock
\urldef\tempurl%
\url{https://api.semanticscholar.org/CorpusID:233987670}
\showURL{%
\tempurl}


\bibitem[Tu et~al\mbox{.}(2024)]%
        {tu2024beyond}
\bibfield{author}{\bibinfo{person}{Haoxin Tu}, \bibinfo{person}{Lingxiao Jiang}, \bibinfo{person}{Debin Gao}, {and} \bibinfo{person}{He Jiang}.} \bibinfo{year}{2024}\natexlab{}.
\newblock \showarticletitle{Beyond a joke: Dead code elimination can delete live code}. In \bibinfo{booktitle}{\emph{Proceedings of the 2024 ACM/IEEE 44th International Conference on Software Engineering: New Ideas and Emerging Results}}. \bibinfo{pages}{32--36}.
\newblock


\bibitem[Utture et~al\mbox{.}(2022)]%
        {Utture2022StrikingAB}
\bibfield{author}{\bibinfo{person}{Akshay Utture}, \bibinfo{person}{Shuyang Liu}, \bibinfo{person}{Christian~Gram Kalhauge}, {and} \bibinfo{person}{Jens Palsberg}.} \bibinfo{year}{2022}\natexlab{}.
\newblock \showarticletitle{Striking a Balance: Pruning False-Positives from Static Call Graphs}.
\newblock \bibinfo{journal}{\emph{2022 IEEE/ACM 44th International Conference on Software Engineering (ICSE)}} (\bibinfo{year}{2022}), \bibinfo{pages}{2043--2055}.
\newblock


\bibitem[Wang et~al\mbox{.}(2024a)]%
        {wang2024makes}
\bibfield{author}{\bibinfo{person}{Haoye Wang}, \bibinfo{person}{Zhipeng Gao}, \bibinfo{person}{Tingting Bi}, \bibinfo{person}{John Grundy}, \bibinfo{person}{Xinyu Wang}, \bibinfo{person}{Minghui Wu}, {and} \bibinfo{person}{Xiaohu Yang}.} \bibinfo{year}{2024}\natexlab{a}.
\newblock \showarticletitle{What makes a good todo comment?}
\newblock \bibinfo{journal}{\emph{ACM Transactions on Software Engineering and Methodology}} \bibinfo{volume}{33}, \bibinfo{number}{6} (\bibinfo{year}{2024}), \bibinfo{pages}{1--30}.
\newblock


\bibitem[Wang et~al\mbox{.}(2024b)]%
        {wang2024just}
\bibfield{author}{\bibinfo{person}{Haoye Wang}, \bibinfo{person}{Zhipeng Gao}, \bibinfo{person}{Xing Hu}, \bibinfo{person}{David Lo}, \bibinfo{person}{John Grundy}, {and} \bibinfo{person}{Xinyu Wang}.} \bibinfo{year}{2024}\natexlab{b}.
\newblock \showarticletitle{Just-in-time todo-missed commits detection}.
\newblock \bibinfo{journal}{\emph{IEEE Transactions on Software Engineering}} \bibinfo{volume}{50}, \bibinfo{number}{11} (\bibinfo{year}{2024}), \bibinfo{pages}{2732--2752}.
\newblock


\bibitem[Wang et~al\mbox{.}(2018)]%
        {wang2018there}
\bibfield{author}{\bibinfo{person}{Junjie Wang}, \bibinfo{person}{Song Wang}, {and} \bibinfo{person}{Qing Wang}.} \bibinfo{year}{2018}\natexlab{}.
\newblock \showarticletitle{Is there a" golden" feature set for static warning identification? an experimental evaluation}. In \bibinfo{booktitle}{\emph{Proceedings of the 12th ACM/IEEE international symposium on empirical software engineering and measurement}}. \bibinfo{pages}{1--10}.
\newblock


\bibitem[Williams and Hollingsworth(2005)]%
        {Williams2005AutomaticMO}
\bibfield{author}{\bibinfo{person}{Chadd~C. Williams} {and} \bibinfo{person}{Jeffrey~K. Hollingsworth}.} \bibinfo{year}{2005}\natexlab{}.
\newblock \showarticletitle{Automatic mining of source code repositories to improve bug finding techniques}.
\newblock \bibinfo{journal}{\emph{IEEE Transactions on Software Engineering}}  \bibinfo{volume}{31} (\bibinfo{year}{2005}), \bibinfo{pages}{466--480}.
\newblock


\bibitem[Xue et~al\mbox{.}(2023)]%
        {xue2023acwrecommender}
\bibfield{author}{\bibinfo{person}{Zhipeng Xue}, \bibinfo{person}{Zhipeng Gao}, \bibinfo{person}{Xing Hu}, {and} \bibinfo{person}{Shanping Li}.} \bibinfo{year}{2023}\natexlab{}.
\newblock \showarticletitle{ACWRecommender: A Tool for Validating Actionable Warnings with Weak Supervision}. In \bibinfo{booktitle}{\emph{2023 38th IEEE/ACM International Conference on Automated Software Engineering (ASE)}}. IEEE, \bibinfo{pages}{1876--1880}.
\newblock


\bibitem[Xue et~al\mbox{.}(2024)]%
        {xue2024selfpico}
\bibfield{author}{\bibinfo{person}{Zhipeng Xue}, \bibinfo{person}{Zhipeng Gao}, \bibinfo{person}{Shaohua Wang}, \bibinfo{person}{Xing Hu}, \bibinfo{person}{Xin Xia}, {and} \bibinfo{person}{Shanping Li}.} \bibinfo{year}{2024}\natexlab{}.
\newblock \showarticletitle{Selfpico: Self-guided partial code execution with llms}. In \bibinfo{booktitle}{\emph{Proceedings of the 33rd ACM SIGSOFT International Symposium on Software Testing and Analysis}}. \bibinfo{pages}{1389--1401}.
\newblock


\bibitem[Xue et~al\mbox{.}(2025)]%
        {xue2025clean}
\bibfield{author}{\bibinfo{person}{Zhipeng Xue}, \bibinfo{person}{Xiaoting Zhang}, \bibinfo{person}{Zhipeng Gao}, \bibinfo{person}{Xing Hu}, \bibinfo{person}{Shan Gao}, \bibinfo{person}{Xin Xia}, {and} \bibinfo{person}{Shanping Li}.} \bibinfo{year}{2025}\natexlab{}.
\newblock \showarticletitle{Clean Code, Better Models: Enhancing LLM Performance with Smell-Cleaned Dataset}.
\newblock \bibinfo{journal}{\emph{arXiv preprint arXiv:2508.11958}} (\bibinfo{year}{2025}).
\newblock


\bibitem[Yan et~al\mbox{.}(2023)]%
        {yan2023closer}
\bibfield{author}{\bibinfo{person}{Dapeng Yan}, \bibinfo{person}{Zhipeng Gao}, {and} \bibinfo{person}{Zhiming Liu}.} \bibinfo{year}{2023}\natexlab{}.
\newblock \showarticletitle{A closer look at different difficulty levels code generation abilities of chatgpt}. In \bibinfo{booktitle}{\emph{2023 38th IEEE/ACM International Conference on Automated Software Engineering (ASE)}}. IEEE, \bibinfo{pages}{1887--1898}.
\newblock


\bibitem[Yang et~al\mbox{.}(2021a)]%
        {yang2021learning}
\bibfield{author}{\bibinfo{person}{Xueqi Yang}, \bibinfo{person}{Jianfeng Chen}, \bibinfo{person}{Rahul Yedida}, \bibinfo{person}{Zhe Yu}, {and} \bibinfo{person}{Tim Menzies}.} \bibinfo{year}{2021}\natexlab{a}.
\newblock \showarticletitle{Learning to recognize actionable static code warnings (is intrinsically easy)}.
\newblock \bibinfo{journal}{\emph{Empirical Software Engineering}}  \bibinfo{volume}{26} (\bibinfo{year}{2021}), \bibinfo{pages}{1--24}.
\newblock


\bibitem[Yang et~al\mbox{.}(2021b)]%
        {Yang2021LearningTR}
\bibfield{author}{\bibinfo{person}{Xueqi Yang}, \bibinfo{person}{Jianfeng Chen}, \bibinfo{person}{Rahul Yedida}, \bibinfo{person}{Zhe Yu}, {and} \bibinfo{person}{Tim Menzies}.} \bibinfo{year}{2021}\natexlab{b}.
\newblock \showarticletitle{Learning to recognize actionable static code warnings (is intrinsically easy)}.
\newblock \bibinfo{journal}{\emph{Empirical Software Engineering}}  \bibinfo{volume}{26} (\bibinfo{year}{2021}).
\newblock


\bibitem[Yang et~al\mbox{.}(2021c)]%
        {yang2021understanding}
\bibfield{author}{\bibinfo{person}{Xueqi Yang}, \bibinfo{person}{Zhe Yu}, \bibinfo{person}{Junjie Wang}, {and} \bibinfo{person}{Tim Menzies}.} \bibinfo{year}{2021}\natexlab{c}.
\newblock \showarticletitle{Understanding static code warnings: An incremental AI approach}.
\newblock \bibinfo{journal}{\emph{Expert Systems with Applications}}  \bibinfo{volume}{167} (\bibinfo{year}{2021}), \bibinfo{pages}{114134}.
\newblock


\bibitem[Yang et~al\mbox{.}(2024)]%
        {yang2024federated}
\bibfield{author}{\bibinfo{person}{Yanming Yang}, \bibinfo{person}{Xing Hu}, \bibinfo{person}{Zhipeng Gao}, \bibinfo{person}{Jinfu Chen}, \bibinfo{person}{Chao Ni}, \bibinfo{person}{Xin Xia}, {and} \bibinfo{person}{David Lo}.} \bibinfo{year}{2024}\natexlab{}.
\newblock \showarticletitle{Federated learning for software engineering: A case study of code clone detection and defect prediction}.
\newblock \bibinfo{journal}{\emph{IEEE Transactions on Software Engineering}} \bibinfo{volume}{50}, \bibinfo{number}{2} (\bibinfo{year}{2024}), \bibinfo{pages}{296--321}.
\newblock


\bibitem[Yuksel and S{\"o}zer(2013)]%
        {Yuksel2013AutomatedCO}
\bibfield{author}{\bibinfo{person}{Ulas Yuksel} {and} \bibinfo{person}{Hasan S{\"o}zer}.} \bibinfo{year}{2013}\natexlab{}.
\newblock \showarticletitle{Automated Classification of Static Code Analysis Alerts: A Case Study}.
\newblock \bibinfo{journal}{\emph{2013 IEEE International Conference on Software Maintenance}} (\bibinfo{year}{2013}), \bibinfo{pages}{532--535}.
\newblock


\bibitem[Y{\"u}ksel and S{\"o}zer(2013)]%
        {yuksel2013automated}
\bibfield{author}{\bibinfo{person}{Ulas Y{\"u}ksel} {and} \bibinfo{person}{Hasan S{\"o}zer}.} \bibinfo{year}{2013}\natexlab{}.
\newblock \showarticletitle{Automated classification of static code analysis alerts: A case study}. In \bibinfo{booktitle}{\emph{2013 IEEE International conference on software maintenance}}. IEEE, \bibinfo{pages}{532--535}.
\newblock


\bibitem[Zhang et~al\mbox{.}(2019)]%
        {Zhang2019AVA}
\bibfield{author}{\bibinfo{person}{Yuwei Zhang}, \bibinfo{person}{Ying Xing}, \bibinfo{person}{Yun zhan Gong}, \bibinfo{person}{Dahai Jin}, \bibinfo{person}{Honghui Li}, {and} \bibinfo{person}{Feng Liu}.} \bibinfo{year}{2019}\natexlab{}.
\newblock \showarticletitle{A variable-level automated defect identification model based on machine learning}.
\newblock \bibinfo{journal}{\emph{Soft Computing}}  \bibinfo{volume}{24} (\bibinfo{year}{2019}), \bibinfo{pages}{1045--1061}.
\newblock


\end{thebibliography}
